\documentclass[journal=jctcce,manuscript=article]{achemso}
\usepackage[utf8]{inputenc}
\usepackage{nicefrac}
\usepackage{amssymb} 
\usepackage{amsmath}
\usepackage{amsthm,bm} 
\usepackage{amsfonts} 
\usepackage{mathtools} 
\usepackage{dsfont}
\usepackage{algorithm}
\usepackage[noend]{algpseudocode}
\usepackage{listings}
\usepackage{booktabs}
\usepackage{siunitx}
\usepackage{multirow}
\usepackage{hyperref}

\newcommand*{\citen}[1]{%
  \begingroup
    \romannumeral-`\x 
    \setcitestyle{numbers}%
    \cite{#1}%
  \endgroup   
}

\def\Tr{\ensuremath\text{Tr}}
\def\br{\ensuremath\bm{r}}

\author{Edoardo Spadetto}
\email{spadetto@scm.com}
\affiliation{Software for Chemistry and Materials NV, NL, 1081HV, Amsterdam, The Netherlands}

\author{Pier Herman Theodoor Philipsen}
\email{philipsen@scm.com}
\affiliation{Software for Chemistry and Materials NV, NL, 1081HV, Amsterdam, The Netherlands}

\author{Arno Förster}
\email{a.t.l.foerster@vu.nl}
\affiliation{Software for Chemistry and Materials NV, NL, 1081HV, Amsterdam, The Netherlands}
\altaffiliation{Theoretical Chemistry, Vrije Universiteit, De Boelelaan 1083, NL-1081 HV, Amsterdam, The Netherlands}

\author{Lucas Visscher}
\altaffiliation{Theoretical Chemistry, Vrije Universiteit, De Boelelaan 1083, NL-1081 HV, Amsterdam, The Netherlands}

\title{Towards Pair Atomic Density Fitting for Correlation Energies with Benchmark Accuracy}

\keywords{RPA, MP2, algorithms, numerical atomic orbitals, non-covalent interactions}

\DeclareUnicodeCharacter{2212}{-}

\begin{document}

\begin{abstract}
Pair atomic density fitting (PADF) has been identified as a promising strategy to reduce the scaling with system size of quantum chemical methods for the calculation of the correlation energy like the direct random phase approximation (RPA) or second-order Møller-Plesset perturbation theory (MP2). PADF can however introduce large errors in correlation energies as the two-electron interaction energy is not guaranteed to be bounded from below. This issue can be partially alleviated by using very large fit sets, but this comes at the price of reduced efficiency and having to deal with near-linear dependencies in the fit set. One posibility is to use global density fitting (DF), but in this work, we introduce an alternative methodology to overcome this problem that preserves the intrinsically favourable scaling of PADF. We first regularize the Fock matrix by projecting out parts of the basis set which gives rise to orbital products that are hard to describe by PADF. After having thus obtained a reliable self-consistent field  solution, we then also apply this projector to the orbital coefficient matrix to improve the precision of PADF-MP2 and PADF-RPA. We systematically assess the accuracy of this new approach in a numerical atomic orbital framework using Slater Type Orbitals (STO) and  correlation consistent Gaussian type basis sets up to quintuple-$\zeta$ quality for systems with more than 200 atoms. For the small and medium systems in the S66 database we show the maximum deviation of PADF-MP2 and PADF-RPA relative correlation energies to DF-MP2 and DF-RPA reference results to be 0.07 and 0.14 kcal/mol respectively. When the new projector method is used, the errors only slightly increase for large molecules and also when moderately sized fit sets are used the resulting errors are well under control. Finally, we demonstrate the computational efficiency of our algorithm by calculating the interaction energies of large, non-covalently bound complexes with more than 1000 atoms and 20000 atomic orbitals at the RPA@PBE/CC-pVTZ level of theory.
\end{abstract}

\maketitle
\mciteErrorOnUnknownfalse

\section{Introduction}

There is great scientific and commercial interest to successfully predict the electronic structure of molecules and materials.
Towards this aim, Density Functional Theory (DFT)\cite{Hohenberg1964,kohn1965,levy79} and the Hartree--Fock (HF) method \cite{Hartree1928,Slater1929,Fock1932,froesefischer87,echenique2007} are indispensable tools, but often they capture exchange-correlation effects only insufficiently. For instance, dispersion and polarization effects which derive mainly from mid- to long-range electron correlation\cite{Stohr2019} are not accounted for by standard DFT functionals.\cite{Burke12,Reilly15}. While empirical dispersion corrections have been highly successful in describing some of these aspects of electron correlation,\cite{Grimme2016, Stohr2019} methods based on coupled cluster (CC)\cite{Coester1958, Coester1960, Cizek1966, Cizek1969, Paldus1972} theory are generally considered to be the most precise class of correlation methods and have been the workhorse for many high precision calculations.\cite{Raghavachari1989} However, given their large computational cost,\cite{bartlett07} their usage is in practice still limited to relatively small systems. Using massively parallel implementations\cite{Gyevi-Nagy2020, Pototschnig2021} and/or local approximations\cite{Neese2009, Riplinger2013, Nagy2017,Nagy2018} it is in principle possible to treat larger molecules.\cite{Al-Hamdani2021, Ballesteros2021} Massively parallel calculations do, however, require large computational resources, while the errors introduced by the local approximations can give rise to uncertainties of the order of several \nicefrac{kcal}{mol} in some cases.\cite{Brandenburg2018,Carter-Fenk2019,Al-Hamdani2021} 

An attractive alternative is provided by double hybrid (DH) density functionals\cite{grimme06, Goerigk2014} which usually offer a good compromise between accuracy and computational effort.\cite{Goerigk2017,Mehta2018,Najibi2018a,Santra2019a} DHs combine DFT with methods which treat correlation effects explicitly, mostly second-order Møller-Plesset perturbation theory (MP2)\cite{moller1934} but (more recently)\cite{Mezei2015,Grimme2016a,Mezei2017} also the random-phase approximation (RPA) \cite{Macke1950, Langreth1975} has been considered for this task. While MP2 is not necessarily very accurate and limited in its applicability, RPA is gaining popularity\cite{Hesselmann2011, Eshuis2011, Eshuis2012a, Ren2012a, Chen2017, Chedid2018, Kreppel2020, Modrzejewski2020,Nguyen2020}. RPA is applicable to a wider class of systems than MP2,  as it is unaffected by some of the drawbacks of MP2, namely divergences for metals, small band gap systems\cite{Macke1950, gruneis10} and large molecules.\cite{ nguyen20}. Without increasing the computational effort one can also greatly improve upon the accuracy of RPA by using $\sigma$-functionals.\cite{Trushin2021, Fauser2021,Erhard2022} Another popular alternative is the inclusion of screened exchange which comes essentially at MP2 cost.\cite{Paier2012, Ren2013, Hummel2019, Forster2022a} 

RPA and MP2 are typically implemented with $\mathcal{O}\left(N^4\right)$\cite{Furche2008} and $\mathcal{O}\left(N^5\right)$ scaling with system size using global density fitting (DF)\cite{Whitten1973,Baerends1973,Dunlap1979,Feyereisen1993} (DF-RPA\cite{Furche2008} and DF-MP2,\cite{Feyereisen1993, Weigend1998} respectively). Efficient implementations of these methods enable applications to systems with a few hundred of atoms even at the quadruple-$\zeta$ level,\cite{Nguyen2020} but larger systems are out of reach on standard hardware. For this reason, more efficient algorithms and approximate implementations have been developed to improve the scaling of both RPA and MP2. Common strategies are the usage of localized orbitals,\cite{Pinski2015,Nagy2016} cluster-in-molecule (CIM) approaches,\cite{Ni2021,Barca2021} or implementations which rely on sparsity in the atomic orbital basis.\cite{Zienau2009,
Doser2009a,
Hohenstein2012,
Hohenstein2012a,
Maurer2014,
Maurer2014a,
Wilhelm2016,
Luenser2017,
Vogler2017,
Beuerle2018,
Graf2018,
Graf2019,
Duchemin2019,
Forster2020,
Glasbrenner2020,
Glasbrenner2021,
Drontschenko2021} In the latter class of methods, implementations using local DF approximations have gained increasing popularity.\cite{Wilhelm2016,Duchemin2019,Forster2020} While they do not achieve linear scaling with systems size, they typically come with a very small prefactor\cite{Forster2020} and are believed to only introduce minor errors compared to canonical, molecular orbital based implementations.\cite{Forster2020,Wilhelm2021} 

One particular flavour of local DF approximations is Pair Atomic Density Fitting (PADF),\cite{Baerends1973,Watson2003,manzer15,Wirz2017,Forster2020} also known also as pair atomic resolution of the identity (PARI),\cite{merlot13,Forster2020} concentric atomic density fitting,\cite{Hollman2014, Hollman2017} or RI-LVL.\cite{Ihrig2015} PADF has originally been introduced to speed up the construction of the Hartree contribution in non-hybrid DFT calculations but was later generalized to accomodate also the formation of the exact exchange matrix in HF or hybrid DFT calculations. For a comparison of PADF-HF to other approximate exact exchange algorithms see ref.~\citen{Rebolini2016}. PADF can also be used to reduce the asymptotic scaling of RPA and Spin-opposite scaled (SOS)-MP2\cite{Jung2004} to formally cubic.\cite{Forster2020} However, quadratic scaling is often observed in practice since the prefactor of the cubic steps is small.\cite{Forster2020} 

This speedup comes at the cost of errors which can in principle cause variational collapse of HF calculations to solutions corresponding to artificially low energies.\cite{Wirz2017} In practice, this is usually only an issue when insufficiently large fit sets are employed. What often is more problematic that PADF also leads to an artificial increase of the magnitude of correlation energies. Unless unrealistically large fit sets are used, this is difficult to avoid these errors and this can then also affect the precision of relative energies that are the typical target of quantum chemical calculations.

In the Amsterdam modelling suite (AMS)\cite{Ruger2022,adf2022}, the issues of PADF are mitigated by applying a projector to the exact HF exchange matrix in order to prevent variational collapse. The same projector can also be applied to orbital coefficients in order to reduce errors in post SCF methods. This strategy has already been successfully employed in the past for many-body perturbation theory (MBPT) based calculations.\cite{Forster2022a,Forster2022c} However, systematic benchmarks against other codes using the same basis sets were not yet performed. For this purpose, we report here an implementation of PADF-MP2 and PADF-RPA in the numerical atomic orbital (NAO) based code BAND.\cite{wiesenekker1988,wiesenekker1991,tevelde91,band22} This implementation allows us to use Gaussian type orbitals (GTO) as basis sets and therefore to systematically investigate the accuracy of the PADF-MP2 and PADF-RPA implementations in AMS for relative correlation energies with respect to global DF based implementations (DF-MP2, DF-RPA) in Psi4 and TURBOMOLE \cite{psi4, Balasubramani2020}. Similar benchmarks of PADF based correlation energies have already been reported by Ihrig et al. using the FHI-AIMS code\cite{Blum2009,Havu2009,Ren2012,FHIaims2009} who found excellent agreement of PADF-MP2 and PADF-RPA with DF-MP2 and DF-RPA\cite{Ihrig2015} and also by Tew.\cite{Tew2018} However, these authors focused on small and medium sized molecules only. To assess whether this accuracy also holds for larger systems and large basis sets, we herein report benchmarks for non-covalently bound dimers with up to 200 atoms and for large GTO-type basis sets up to quintuple-$\zeta$ (5Z). In our benchmarks, we focus exclusively on non-covalent interactions. This is mostly due to the availability of accurate reference values for datasets containing large molecules, like the L7\cite{Sedlak2013} or the S30L\cite{Sure2015} compilations.

This paper is organized as follows: In section~\ref{sec::theo} we review the PADF method and introduce the projector method (PM). We also sketch how PADF can be used to achieve low-order scaling implementations of RPA and SOS-MP2. For more details, we refer to previous work.\cite{Forster2020,Forster2020b}
After an outline of our computational details in section~\ref{sec::comp}, we assess the accuracy of relative PADF-MP2 and PADF-RPA correlation energies in section~\ref{sec::res}. Our calculations show that PADF-SOS-MP2 is in excellent agreement to DF-SOS-MP2. Our interaction energies for the S66 dataset\cite{Rezac2011} show maximum absolute deviations for PADF-MP2 and PADF-RPA with respect to the reference results of 0.07 and 0.14 kcal/mol respectively, irrespective of the chosen basis set. For much larger molecules, we observe only a negligible loss in accuracy for SOS-MP2 and we find the PM to be decisive to obtain accurate results. The loss in accuracy is more pronounced for PADF-RPA, but errors are smaller than errors due to basis set incompleteness or due to local correlation approximations for large molecules\cite{Ballesteros2021,Al-Hamdani2021} To showcase the efficiency of our implementation, we calculate PADF-RPA interaction energies of eight large non-covalently bound complexes at the triple-$\zeta$ (TZ) level, with up to 1000 atoms and more than 20000 AOs. Finally, section~\ref{sec::sum} summarizes and concludes this work.
\section{\label{sec::theo}Theory}
Throughout this paper, the indicies $\{i,j,...\}$ ($\{a,b,...\}$) refer to occupied (virtual) orbitals, and the indices $\{p,q,...\}$ refer to general molecular orbitals. Primary basis functions are labeled with $\{\mu,\nu,\kappa,\lambda,...\}$ while $\{\alpha,\beta,\gamma,...\} $ denote fit functions. $\{A, B, C, ...\}$ denote atoms. $o$ is a generic index which can either denote a primary basis function or a fit function.

\subsection{Density Fitting}
We use Mulliken notation throughout this work, in which the generic form of two-electron integrals is given by
\begin{equation}
\label{integral-kernel}
    \mathcal{K}_{\mu \nu \kappa \lambda} = 
    \int d \br d\br' 
    \chi^*_{\mu}(\br) 
    \chi_{\nu}(\br) 
    \mathcal{K}(\br,\br') 
    \chi^*_{\kappa}(\br') 
    \chi_{\lambda}(\br') \;,
\end{equation}
with $\mathcal{K}(\br,\br')$ being a general (non-local) kernel. Important examples for $\mathcal{K}$ are the electron-electron interaction, $v_c(\br,\br') = \frac{1}{|\br - \br'|}$ which is a key ingredient in HF and post-HF methods, and the non-interacting polarizability $P(\br,\br')$ (for a certain value of imaginary frequency or time) which appears for instance in RPA or SOS-MP2. The symbol $\chi_{\mu}$ refers to an atom-centered basis function which belongs to a basis set of $N_\text{bas}$ functions $\{\chi_{\mu}(\textbf{r}) \in X \hspace{4pt}\forall \hspace{4pt} 1 \le \mu \le N_\text{bas}  \}$. We assume these functions to be composed of an angular part $Y_{l}^{m}(\theta, \phi)$ and a radial function $R_{n}(|\textbf{r}_{A}| )$,
\begin{equation}
\chi_{\mu}(\br) = \chi_{lmn, A}(\theta,\phi,|\textbf{r}_{A}|) =  Y_{l}^{m}(\theta, \phi) R_{n}(|\textbf{r}_{A}| ) \;.
\label{eq:basfun}
\end{equation}
The radial part only depends on the distance from atom $A$, $\textbf{r}_{A}$. The angular part $Y_{l}^{m}$ is a spherical harmonic function with angles defined in the local coordinate system of atom A.

Representing $\mathcal{K}$ in this way, the memory required to store all the integrals defined by \eqref{integral-kernel} grows as $\mathcal{O}\left(N^4\right)$ with system size. Furthermore, evaluating these electron-electron interaction integral explicitly is difficult, when sets of STOs or NAOs are chosen as primary basis. It is therefore convenient to look in more detail at the set of functions $F_{p} =  \{\chi_{\mu}(\br)\chi_{\nu}(\br)  \hspace{4pt}\forall \hspace{4pt} 1 \le \mu \le N_\text{bas} \text{ and } 1 \le \nu \le N_\text{bas}\}$ that is obtained by gathering all unique products of two basis functions and investigate whether this function set can be represented in a more compact form via density fitting. 

We first define $\hat{\mathcal{K}}$ formally\cite{Loaiza2017} as a linear operator in a Hilbert space $\mathcal{H^K}$: $\hat{\mathcal{K}} f(\br) = \int d \br'\mathcal{K}(\br,\br') f(\br')$, with the inner product on $\mathcal{H^K}$ defined as
\begin{align}
(f |g) = \int f(\br) \; \hat{\mathcal{K}} \; g(\br) d\br.
\label{eq:scalarproduct}
\end{align}
The set of fit functions that is defined in (PA)DF forms the basis $F_{f}$ and spans a subspace $X_f$ of the full Hilbert space $\mathcal{H^K}$. Given $f_{\alpha}(\br)$ and $f_{\beta}(\br) \in F_f$,  the expansion of $\hat{\mathcal{K}}$ in $F_f$ is

\begin{equation}
\label{integral-kernel2}
    \mathcal{K}_{\alpha\beta} = (f_\alpha |f_\beta).
\end{equation}

Likewise we have that the basis $F_{p}$ spans the space $X_{p} \subset \mathcal{H^K}$, with the integrals taking the form \eqref{integral-kernel} or more concisely $(\chi_{\kappa}\chi_{\lambda} |\chi_{\mu}\chi_{\nu})$. If $X_{p} \subseteq X_{f}$, $\chi_{\mu}(\br) \chi_{\nu}(\br)$ can be expressed exactly in terms of fit functions  and we can consequently use the compact expression \eqref{integral-kernel2} instead of \eqref{integral-kernel}. In practice we find that part of the product space is not spanned by the fit functions ($X_{p} \setminus X_{f} \neq \emptyset$). To characterize errors made by fitting basis functions products with the fit set $F_{f}$ we therefore write members of the product basis $F_{p}$ as
\begin{align}
    \chi_{\mu}(\br)\chi_{\nu}(\br) = \sum_{\alpha} f_{\alpha}(\br) c_{\mu\nu\alpha}  + e_{\mu\nu}(\br)
    \label{eq:expansions}
\end{align}
where $e_{\mu \nu}(\br)$ is an error function which accounts for the fact that $X_f$ does not completely span $X_{p}$. To keep the notation short we only indicate explicitly the dependence of this error function of the basis function pair indices $\mu$ and $\nu$ and omit the dependencies on the choice of fit set and the optimization criterion used to determine the fit coefficients $c_{\mu\nu\alpha}$. The exact representation of $\hat{\mathcal{K}}$ in $X_{p}$ can then be written as
\begin{equation}
    \label{eq:expansions_k}
\begin{aligned}
    \mathcal{K}_{\mu\nu\kappa\lambda} =  \sum_{\alpha\beta} c_{\mu\nu\alpha}\mathcal{K}_{\alpha\beta}c_{\kappa\lambda\beta} +  &  \sum_{\alpha}c_{\mu\nu\alpha} \int d\br  f_{\alpha}(\br)\hat{\mathcal{K}}e_{\kappa\lambda}(\br)  + \\ &  \sum_{\beta} c_{\kappa\lambda\beta} \int d\br e_{\mu\nu}(\br)\hat{\mathcal{K}}f_{\beta}(\br) + \int d\br e_{\mu\nu}(\br)\hat{\mathcal{K}}e_{\kappa\lambda}(\br).
\end{aligned}
\end{equation}

Using the scalar product notation  \eqref{eq:scalarproduct} we also have
\begin{align}
     ( f_{\alpha} | \chi_{\mu}\chi_{\nu} ) &  = \sum_{\beta} \mathcal{K}_{\alpha\beta}  c_{\mu\nu\beta}  + ( f_{\alpha} |e_{\mu\nu})& 
     \label{eq:coefficients}
\end{align}

Keeping in mind that the function $e_{\mu\nu}$ depends implicitly on the fit space and on the kernel used to define the scalar product, we can define this function to lie entirely in $X_{e} = X_{p} \setminus X_{f}$ so that we have:
\begin{align}
    ( f_{\alpha} |e_{\mu\nu}) =0 .
    \label{eq:definition}
\end{align}
Note that integrals over $f_{\alpha}$ and $e_{\mu\nu}$ with other kernels are in general non-zero. Then, assuming $\mathcal{K}_{\alpha\beta}$ invertible and considering the symmetry of scalar products,we may write 
\begin{equation}
    c_{\mu\nu\alpha}= \sum_{\beta} (  \chi_{\mu}\chi_{\nu} | f_{\beta} ) (\mathcal{K}^{-1})_{\beta\alpha} .
    \label{eq:fitcoeff}
\end{equation}

This choice of fit coefficients in \eqref{eq:fitcoeff} can also be viewed as minimizing the Lagrangian
\begin{align}
    \label{eq:lagrangian}
   \mathcal{L}_{\mu\nu} & = (  e_{\mu\nu} \vert  e_{\mu\nu} )  \\ 
   \nonumber
   & = (  {\chi_{\mu}\chi_{\nu}} \vert  \chi_{\mu}\chi_{\nu} ) - 2 \sum_{\beta }c_{\mu\nu\beta}  (\chi_{\mu}\chi_{\nu} | f_{\beta}) +  \sum_{\alpha\beta} c_{\mu\nu\alpha}c_{\mu\nu\beta} ( f_{\alpha} | f_{\beta})
\end{align}

for every $e_{\mu\nu}$. Minimizing $\mathcal{L}_{\mu\nu}$ guarantees a reasonable precision also for off-diagonal terms in the residuals matrix since due to the Cauchy-Schwartz inequality we have
\begin{align}
     (  e_{\mu\nu} \vert  e_{\kappa\lambda} ) \leq \sqrt {\mathcal{L}_{\mu\nu}  \mathcal{L}_{\kappa\lambda} } \;.
     \label{eq:cs}
\end{align}

With the fit coefficient definition \eqref{eq:definition}, the cross error terms in \eqref{eq:expansions_k} vanish. This property  arises naturally because the metric used in fitting is defined through the same kernel we aim to fit. 
Using $\mathcal{K} = v_c$ and functions of the form \eqref{eq:basfun}, the theoretical framework just presented is known as (global) robust Density Fitting (DF).\cite{ Whitten1973, Baerends1973, Dunlap1979, Dunlap1979a, Dunlap1983a, Feyereisen1993} 
It has the advantage of reducing the storage complexity of the matrix elements of the kernel and the amount of integrals to be evaluated from $\mathcal{O}\left(N^4\right)$ to $\mathcal{O}\left(N^3\right)$. DF is not an approximation if the expansion is complete, and in this case a compression would only be achieved for exact linear dependencies in $X_{p}$. In practice the compression is obtained at the price of an approximation since for reasons of computational efficiency $F_f$ does not span the complete space of products of primary basis functions. Considering what is left out, we note that a product set defined as $\chi_{\mu}\chi_{\nu}$ is strongly non-orthogonal. Orthogonalization of such a basis to span as much as possible of the full Hilbert space would result in linear combinations of $\chi_{\mu}\chi_{\nu}$ with large coefficients. Given the finite precision of computer operations, the calculation of matrix representations of these parts of $\mathcal{H^K}$ is likely to be numerically unstable. In addition, we can expect such combinations of pair functions to be of minor physical relevance for a quantum chemical calculation. For this reason it is also numerically favourable to work with a fit set that is better behaved in terms of orthogonality than an orthogonalized product set. 

DF reduces the asymptotic scaling of the evaluation of RPA and direct MP2 correlation energies from $\mathcal{O}\left(N^6\right)$ and $\mathcal{O}\left(N^5\right)$, respectively to $\mathcal{O}\left(N^4\right)$.\cite{Furche2008} However, the asymptotic scaling of methods involving exchange terms is not automatically reduced. For instance, using \eqref{eq:expansions_k}, the exchange contribution to the Fock matrix can be expressed as\cite{Weigend2002a} 
\begin{equation}
    F_{\mu \nu} = \sum_{\kappa \lambda}  P_{\kappa \lambda}   \sum_{\alpha \beta} c_{\mu \kappa \alpha} 
    v_{\alpha \beta} 
    c_{\nu \lambda \beta} \;,
\end{equation}
which is evaluated with the same asymptotic scaling of $\mathcal{O}\left(N^4\right)$ as the variant without density fitting when no further approximations are made. This is also generally true for post-HF methods where exchange terms profit less from the compression of 4-index tensors.\cite{Hohenstein2012, Forster2020} Therefore, it is advantageous to eliminate the exchange terms entirely and introduce empirical scaling factors, as for instance in the SOS-MP2\cite{Jung2004} or SOS-CC2 methods.\cite{Winter2011} 
 
One can however also greatly improve upon the efficiency of global DF by constraining eq.~\eqref{eq:expansions} in such a way that the number of non-zero-elements in $c$ only grows linearly with system size. For instance, instead of using the Coulomb kernel directly\cite{Whitten1973, Dunlap1979} one can avoid to define the scalar product on the operator $\hat{\mathcal{K}}$ and define another metric with more suitable properties. This could for instance be a local kernel, like the overlap kernel\cite{Dunlap1979} (also known as RI-SVS) or an attenuated Coulomb kernel.\cite{Feyereisen1993, Jung2005} These kernels have been used successfully to lower the complexity of for instance $GW$\cite{Wilhelm2018,Wilhelm2021}, RPA\cite{Wilhelm2016,Luenser2017}, MP2\cite{Glasbrenner2020,Glasbrenner2021} and CC2\cite{Sacchetta2022} calculations. The price to be paid is the loss of robustness, equation \eqref{eq:definition} is then not fulfilled so that the cross terms in \eqref{eq:expansions_k} become non-zero. An alternative approach which introduces the desired sparsity in the fit-coefficient tensor more directly is PADF. In PADF the density fit is restricted to pairwise sums only and subsequently distance cut-offs are introduced. Using Latin uppercase superscripts to denote the atomic centers of functions, the PADF expansion of products of basis functions is
\begin{align}
\label{eq:padf-expansion}
    \chi^{A}_{\mu}(\br)\chi^{B}_{\nu}(\br) = 
    \sum_{\alpha \in A}c_{\nu\mu\alpha} f_{\alpha}(\br) + \sum_{\beta \in B}c_{\mu\nu\beta}f_{\beta}(\br) + e_{\mu\nu}(\br) \;,
\end{align}
replacing the simpler expansion \eqref{eq:expansions}. The notation $\alpha \in A$ indicates that the summation is restricted to fit functions centered on atom $A$. 

\subsection{Fit Set Generation}
\label{sec:fitsetgeneration}
\subsubsection{Fit sets from products of basis functions}
It is easily understood that the choice of $F$ is of key importance in a PADF code. Ideally, the fit set should be generated on-the-fly, tailored to the primary basis at hand and the precision of the expansion \eqref{eq:padf-expansion} should be adjustable in a systematic way using only a single parameter. Many algorithms for this task have been developed for global DF.\cite{Kallay2014, Stoychev2017, Lehtola2021, Semidalas2022} An alternative way to generate fit sets on-the-fly is Cholesky decomposition,\cite{Koch2003, Aquilante2007, Aquilante2009} but this approach is not straightforwardly generalized to codes which can not evaluate 3-center integrals involving the Coulomb potential analytically. We are only aware of two algorithms which have been developed specifically for PADF.\cite{Ihrig2015, Medves2022} We here adopt the one by Ihring and coworkers\cite{Ihrig2015} which we recapitulate for completeness. 

From all unique combinations of AOs centered on atom $A$ (denoted by $X_{A}$) we build an atom-specific trial fit set $\tilde{F}_{A}$ of functions of form \eqref{eq:basfun}, 
\begin{align}
\tilde{F}_{A} = \{\tilde{f}_{\mu\nu}(\textbf{r})  \text{ such that } & R_{\mu\nu}(|\textbf{r}_{A}|)= R_{\mu}(|\textbf{r}_{A}|) R_{\nu}(|\textbf{r}_{A}|) \text{ and }   \\  & | l_{\mu}-l_{\nu} |< l_{\mu\nu} < | l_{\mu}+l_{\nu} |  \text{ }  \forall \text{ } \chi_{\mu}(\textbf{r}) \text{ and } \chi_{\nu}(\textbf{r}) \in X_{A}  \} \nonumber
\end{align}
We then regroup $F_{A}$ in subsets with same angular momentum 
\begin{align}
    \tilde{F_{A}} = \bigcup_{l,m} \tilde{F}_{A,lm} =  \bigcup_{l,m} {Y}_{l}^{m}(\theta,\phi) \times \tilde{R}_{A,lm}
\end{align}
where $\tilde{R}_{A,lm}$, the set of radial components centered on the same atom $A$, is multiplied element-wise to the same spherical harmonic. We then compute the eigenvectors of the matrix $(\tilde{R}_{\alpha}|\tilde{R}_{\beta})$ for  $\tilde{R}_{\alpha} \text{ and } \tilde{R}_{\beta} \in \tilde{R}_{A,lm}$ and we keep only the ones with eigenvalue greater than a specific threshold $\epsilon_\text{fit}$. This threshold can be seen as a parameter tuning the fit quality. Setting it to zero does not imply exact fitting, as it only solves the one-center part exactly. In addition, as mentioned above, choosing a too small parameter will likely introduce numerical instabilities. 
The set of remaining eigenvectors is called $R_{A,lm}$. Our final fit set is then
\begin{align}
    F = \bigcup_{Alm} F_{A,lm} =  \bigcup_{Alm} Y_{l}^{m}(\theta,\phi) \times R_{A,lm}
    \label{eq::trialfit}
\end{align}
Alg.~\ref{algo:fit} shows a pseudocode of the algorithm. In alg.~\ref{algo:fit}, $R_{\mu}(r) \times \{Y^{m}_{l_u}(\theta,\phi)\}$  denotes basis functions with same radial part and different spherical harmonic.

\makeatletter
    \def\BState{\State\hskip-\ALG@thistlm}
    \makeatother

    \begin{algorithm}
    \setstretch{1.5}
        \small
        \caption{\texttt{ Fit set generation algorithm }}\label{euclid}
        \begin{algorithmic}[1]
            \For{ \texttt{every atom $A$} }
            \For{ \texttt{every $R_{\mu}(r) \{Y^{m}_{l_u}(\theta,\phi)\}$ in $X_{A}$}}
            \For{ \texttt{every $R_{\nu}(r) \{Y^{m}_{l_v}(\theta,\phi)\}$ in  $X_{A}$}}
        
            \State$\tilde{F}_{A} \texttt{ push back } R_{\mu}(r)R_{\nu}(r) \{Y^{-l}_{l= |l_{u}-l_{v}|}, ... , Y^{+l}_{l= |l_{u}+l_{v}|}\}$
            \EndFor
            \EndFor
            \For{ \texttt{every $\tilde{F}_{A,lm} $ in $\tilde{F}_{A}$ } }
            \For{ \texttt{every couple $\tilde{f}_{a}, \tilde{f}_{b}$ in $\tilde{F}_{A,lm} $ } }
            \State  $V_{ab} = ( \tilde{f}_{a}(\textbf{r}) |  \tilde f_{b}(\textbf{r}) )$ 
            \EndFor
         
            \State \texttt{Find Spectrum of $V \Rightarrow  V_{ab} =  v_{a\alpha} \varepsilon_{\alpha\beta} v^\dagger_{\beta b}$} \texttt{(eigenvalues in increasing order)}
          
            \State \texttt{$P_{\alpha b}= \sqrt{\varepsilon_{\alpha\beta}} v^{\dagger}_{\beta b} $ (  $m$ s.t. $\varepsilon_{\alpha\alpha}>\epsilon_{a} $ $\Rightarrow \alpha = 1 .. m$ )}
            \State \texttt{$\tilde{F}_{A,lm}  = P \cdot \tilde{F}_{A,lm}   $} \texttt{(sets assumed to be matrices, same ordering used in V)}
            \EndFor

            \EndFor
         
        \end{algorithmic}
       \label{algo:fit}
    \end{algorithm}
For small basis sets, the fit set generated through such a procedure does sometimes not lead to sufficiently accurate results. 
To overcome this problem, Ihrig et al.\cite{Ihrig2015} artificially enlarged the fit set. This is achieved by adding a new function to each $X_{A}$.  In our implementation, the function is a Slater type orbital with angular momentum $l^A_{max} + 1$ where $l^A_{max}$ is the maximum angular momentum present in $X_{A}$, and the arbitrarily chosen exponent $\alpha$ equal to its angular momentum. 
\begin{align}
\chi(\textbf{r}) = r^{\alpha} \exp(- \alpha r) Y^{}_{\alpha}(\theta,\phi) && \alpha = l^A_{max}+1
\end{align}
In the following, we will refer to this procedure as $L$-enhancement ($L$-e).

\subsubsection{Slater type orbital fit sets}
As an alternative to the algorithm just described, we herein also test the use of hand-optimized STO type fit sets. These come with the disadvantage that they are not systematically improvable through the adjustment of a single parameter $\epsilon_{\text{fit}}$. They are however more compact and therefore more suitable for large-scale applications. In this work we use three different thresholds which we refer to as \emph{Normal}, \emph{Good} and \emph{VeryGood}. The former two contain STO type functions with angular momentum up to $l=4$, while the latter one contains functions with exponents up to $l=6$. We have described these fit sets in ref.~\citen{Forster2020} to which we refer for more details. 

\subsection{Projection Methods}
The improved algorithmic scaling deriving from PADF comes with downsides which needs to be handled carefully in order to retain sufficient numerical precision. Therefore, we use two related projection methods which we describe in the following:

\subsubsection{Projection Method for the Basis Set}
To prevent instabilities due to near-linear dependencies in the AO basis set itself, this basis set size is often reduced by modifying the L\"{o}wdin orthornomalization\cite{Lowdin1970, Lehtola2020} step. The L{\"o}wdin transformation matrix $\mathbf{S}^{-1/2}$ to an orthonormal set follows from the eigensystem of the overlap matrix
\begin{align}
    \mathbf{S} = & \mathbf{U} \mathbf{D} \mathbf{U}^T \\
    \mathbf{S}^{-1/2} = & \mathbf{U} \mathbf{D}^{-1/2} \mathbf{U}^T \;.
\end{align}
Here we can choose to ignore eigenvectors with eigenvalues below a threshold $\epsilon_\text{bas}$. The simplest way to achieve this is to set the corresponding eigenvectors in $U$ to zero (thus introducing artificial states). This is done in the ADF implementation. In the BAND implementation, we define a smaller orthonormal basis by introducing the regularized L{\"o}wdin  transformation
\begin{equation}
\tilde{\mathbf{S}}^{-1/2} = \mathbf{S}^{-1/2} \tilde{\mathbf{U}}
\label{eq:lowdindepfix}
\end{equation}
where $\tilde{\mathbf{U}}$ is the non square (tall) matrix obtained by removing the eigenvectors columns with eigenvalues smaller than $\epsilon_\text{bas}$ from $\mathbf{U}$. Doing so, fewer orbitals are obtained and appearance of artificial states is avoided. The elimination of problematic orbitals that are expressed in the original basis with large coefficients also helps to prevent problems later on when considering the product basis and can therefore be beneficial to mitigate errors resulting from the density fitting.

\subsubsection{HF projection method}
A known issue of PADF is that contributions to the electron repulsion energies can become negative which can lead to variational collapses of HF SCF calculations\cite{merlot13,Wirz17}. As we will argue below, this problem is to a large extent due to integrals over product functions that are difficult to describe by the fit set. A way to avoid this problem would be to Cholesky decompose the matrix \eqref{integral-kernel} and keep only the most important Cholesky vectors, but this is not practical in calculations with Slater or numerical type orbitals. Instead, the PADF implementation in AMS uses a simple projector in the original AO space
\begin{align}
     \mathbf{T} =  \mathbf{T} \mathbf{R} \mathbf{T}^T
     \label{eq:proj}
\end{align}
where we use the eigensystem of the AO overlap matrix
\begin{align}
     \mathbf{S} =  \mathbf{U} \mathbf{D} \mathbf{U}^T
    \label{eq:preproj}
\end{align}
with $\mathbf{D}$ being a diagonal matrix with the eigenvalues on the diagonal, and $\mathbf{U}$ having the eigenvectors stored as columns.
The diagonal matrix $\mathbf{R}$ is obtained from $\mathbf{D}$ by taking
\begin{align}
    R_{ij} = \delta_{ij} \Theta (D_{ii}-\epsilon_{K})  
\end{align}
Note that for $\epsilon_{K}=0$ we get $\mathbf{R}=\mathbf{1}$ and hence $\mathbf{T}=\mathbf{1}$. Heuristically, the action of $\mathbf{T}$ on a vector or matrix is to remove components of the space parallel to eigenvectors of the overlap matrix with eigenvalues smaller than the specified threshold $\epsilon_{K}$. At the SCF stage, the projector is applied both on the left and on the right of the  exact exchange matrix $\mathbf{K}$ 
\begin{align}
    \tilde{\mathbf{K}} = \mathbf{T}^T    \mathbf{K}    \mathbf{T}  
    \label{eq:Kmatproj}
\end{align}
While the regularized L\"{o}wdin orthonormalization removes a subspace for all energy terms, the HF projector method neglects only the (small) stabilizing action of the exchange energy, shifting energies upwards. The default value for $\epsilon_K=10^{-3}$ in BAND can therefore be much higher than the one for the L\"{o}wdin orthonormalization projector $\epsilon_\text{bas}=10^{-8}$.

We also use the same projector (\ref{eq:proj}) to calculate correlation energies. For this we redefine the matrix $\mathbf{b}$ which transforms from the AO to the MO basis as
 \begin{equation}
 \label{eq:transformed-coefficients}
     b'_{i\mu} = \sum_{\mu'} b_{i\mu'} T_{\mu'\mu} \;.
 \end{equation}
Here, the use of the projector is supposed to improve the accuracy of correlation energies by removing a subspace leading to AO products which can only be represented poorly by the fit set. The usefulness of the application of the PM-$K$ to correlation energies can be rationalized as follows: If we consider an eigenvector $v_{o}(\textbf{r})= \sum_{\mu}s_{o\mu}\chi_{\mu}(\textbf{r}) $ of the overlap matrix, we can notice that $ \sum_{o} s^\dagger_{\mu o} s_{ o\nu} = D^{-1}_{\mu\nu}$. From this we understand that the average order of magnitude of the coefficients is at least of the order of $s_{o\mu} \sim \frac{1}{\sqrt{D_{\mu\mu}}}$. This shows that the basis set poorly describes eigenvectors relative to small eigenvalues of $S$, since large coefficients are needed to expand a small orthogonal component. We then expect that products involving such linear combinations are the most difficult to fit. This is mostly because of the diffuse products of basis functions from distant atoms and from the consequent difficulties in using the PADF approximation to express such products.\cite{Forster2020b} 

\subsection{RPA and SOS-MP2 correlation energies}

We now briefly discuss how PADF can be used to speed up the evaluation of SOS-MP2 and RPA correlation energies, summarizing the more detailed discussions in ref. \citenum{Forster2020} and \citenum{Forster2020b}. In the basis of fit functions, the RPA correlation energy can be expressed as 
\begin{equation}
    \label{e_rpa_comp}
    \begin{aligned}
    E^\text{RPA}_c = & \frac{1}{2\pi}
    \int^{\infty}_0  d \omega \Tr
    \left\{ \left[\log\left(\mathbf{1} - \mathbf{Z} (i\omega)\right)\right] + 
    \mathbf{Z} (i\omega)\right\} \\ 
    = & \sum^{N_{\omega}}_{k} \sigma_k \Tr
    \left\{ \left[\log\left(\mathbf{1} - \mathbf{Z} (i\omega_k)\right)\right] + 
    \mathbf{Z}  (i\omega_k)\right\} \;,
    \end{aligned} 
\end{equation}
which follows directly from the corresponding real-space representation of the RPA.\cite{Langreth1977} The integration is performed over the imaginary frequency axis for which either modified Gauss-Legendre grids or, more efficiently,\cite{Kaltak2014} minimax grids of size $N_{\omega}$ can be used. $\omega_k$, $\sigma_k$ denote points and corresponding integration weights on the imaginary axis. $\mathbf{Z}$ is
\begin{equation}
Z_{\alpha\beta}(i\omega) = \sum_{\gamma} P^{(0)}_{\alpha\gamma,i\omega}v_{\gamma\beta} \;,
\label{eq:Z}
\end{equation}
and is obtained through the non-interacting polarizability $P^{(0)}$ and the electron-electron interaction $v$ in the basis of fit functions. Since matrix logarithms are difficult to calculate, we use that (assuming $\mathbf{1} -\mathbf{Z}$ can be diagonalized with eigenvalues $\lambda_j$)
\[
\Tr \left[ \log ( \mathbf{ 1} - \mathbf{Z} )  \right] = \sum_{j} \log(\lambda_j) = \log \left(\prod_j \lambda_j \right) = \log \vert   \mathbf{ 1} - \mathbf{Z}  \vert \;,
\]
and evaluate the determinant $\vert \mathbf{1} - \mathbf{Z} \vert $ instead. The imaginary frequency representation of $P^{(0)}$ is obtained from its discrete imaginary time representation using nodes $\left\{\tau_k\right\}_{j = 1, \dots , N_{\tau}}$. The transformation is achieved by the discrete cosine transform (since $P^{(0)}$ is bosonic) 
\begin{equation}
 \label{exactFFT}
 P^{(0)}(i\omega_k) = 
 -i\sum^{N_{\tau}}_{j}\gamma^{(c)}_{kj}\cos (\omega_k \tau_j) P^{(0)}(i\tau_j) \;,
 \end{equation}
 where the $\gamma^{(c)}_{kj}$ are the matrix elements of the kernel of the discrete cosine transform. The imaginary time representation of $P^{(0)}$ is given as
 \begin{equation}
 \label{rpa_sparse}
 P^{(0)}_{\alpha \beta}(i\tau_j)  = 
-i c_{\mu \nu \alpha}G^{(0),<}_{\mu \kappa} (-i\tau_j) 
G^{(0),>}_{\nu \lambda} (i\tau_j) c_{\kappa \lambda \beta} \;,
 \end{equation}
 with greater and lesser components of the time-ordered Green's functions being defined as
 \begin{align}
\label{greensGWbasisO}
    G^{(0),<}_{\mu \nu}(i\tau_j) = & -i\sum_i
    b'_{\mu i } e^{-\epsilon_i\tau_j}b'_{i \nu}\\
\label{greensGWbasisU}
    G^{(0),>}_{\mu \nu}(i\tau_j) = & -i\sum_a
    b'_{\mu a } e^{-\epsilon_a \tau_j}b'_{a \nu} \;.
\end{align}
Notice, that $b'$, as defined in \eqref{eq:transformed-coefficients}, appears in these equations so that the the PM enters the PRA correlation energies through $G^{\lessgtr}$. As previously discussed in some detail by us,\cite{Forster2020b} the evaluation of \eqref{rpa_sparse} scales asymptotically as $\mathcal{O}\left(N^2\right)$ with system size in the PADF approximation, since the number of elements in $c$ only scales as $\mathcal{O}\left(N\right)$ with system size. For detailed working equations we refer to our previous work.\cite{Forster2020,Forster2020b} In practice, the efficiency of this approach depends on the possibility to represent the imaginary time and frequency dependencies of the polarizability as compactly as possible. We follow Kaltak and Kresse\cite{Kaltak2014,Kaltak2014a} and use non-uniformly spaced minimax and least square grids as described in ref.~\citen{Helmich-Paris2016} for imaginary time, and ref.~\citen{Forster2021} for the imaginary frequency domain.

Alternatively, the imaginary frequency polarizability $ P^{(0)}(i\omega_k)$ can be evaluated directly in the MO basis as for instance described in ref.~\citen{Ren2012},
\begin{equation}
\label{rpa_canonical}
    P^{(0)}_{\alpha\beta}(i\omega_k) = -\sum_{a}\sum_i c_{ia \alpha} \frac{1}{\epsilon_a - \epsilon_i - i\omega_k}c_{ia \beta} + c.c.
\end{equation}
where
\begin{equation}
    c_{ia \alpha} = \sum_{\mu \nu} b'_{\mu i } c_{\mu \nu \alpha} b'_{\nu a } \;.
    \label{eq:projectorrpa}
\end{equation}

Using the series expansion of the logarithm in \eqref{e_rpa_comp} one obtains the direct term of the MP2 correlation energy, $E_c^{(2)}$ as its second-order term in $v_c$. $E_c^{(2)}$ can be evaluated directly in imaginary time and is given by 
\begin{equation}
\label{e_mp2_direct}
\begin{aligned}
E^{(2)}_{c} = & -\frac{1}{2}\sum^{N_{\tau}}_{k} \alpha_k \Tr \left( \sum_{\gamma}Z_{\alpha \gamma, \tau_k} Z_{\gamma \beta, \tau_k}\right) \\
= & -\frac{1}{2}\sum^{N_{\tau}}_{k} \alpha_k \sum_{\alpha\beta} 
Z_{\alpha \beta, \tau_k} Z_{\beta \alpha, \tau_k} \;,
\end{aligned}
\end{equation}
where $\left\{\alpha_k\right\}_{k = 1, N_{\tau}}$ denote the integration weights corresponding to the points $\left\{\tau_k\right\}_{k = 1, N_{\tau}}$.
In the spin-polarized case we need to calculate
\begin{equation}
\begin{aligned}
Z_{\alpha \beta, \tau_k} Z_{\beta \alpha, \tau_k} = &
\sum_{\sigma = \alpha,\beta}
\sum_{\sigma' = \alpha,\beta}
Z_{\alpha \beta, \sigma, \tau_k}
Z_{\alpha \beta, \sigma', \tau_k} \;.
\end{aligned}
\end{equation}
When working in the AO basis, we are only interested in the contribution to $E_c^{(2)}$ from electrons with unpaired spins which is used for instance in spin-opposite scaled (SOS) MP2\cite{Jung2004} or in DOD-DHs.\cite{Kozuch2011} In that case, only the terms with $\sigma \neq \sigma'$ contribute and the resulting correlation energy expression is scaled by an empirical factor.

While PADF can also be used to lower the time complexity of the exchange contribution to MP2 from $\mathcal{O}\left(N^5\right)$ to $\mathcal{O}\left(N^3\right)$\cite{Forster2020}, the resulting working equations can only be implemented with a very high prefactor and are therefore not useful in practice. Instead, the full MP2 correlation energy is evaluated in the MO basis
\begin{align}
E_\text{MP2} = \sum_{ijab}\frac{ (  ia \vert jb )^2 +\frac{1}{2} [ (  ia \vert jb ) - (  ib \vert ja ) ]^2 }{\varepsilon_{i}+\varepsilon_{j}-\varepsilon_{a}-\varepsilon_{b}} \;.
\label{eq:MP2}
\end{align}
The expression is evaluated as in typical DF-MP2 codes.\cite{Cooper2017} The necessary fit-coefficients are again transformed to the MO basis using the relation \eqref{eq:projectorrpa}.

\subsection{Summary of the projection method thresholds}
To improve the numerical accuracy we apply at various stages of the calculation related techniques which we loosely call projector methods. Although the eigenvectors of the overlap matrix form an orthogonal basis set, some of the functions thus constructed are in an Orwellian sense more equal than others: those with a small eigenvalue are less valuable, and may even be numerically harmful.
In this spirit the first threshold $\epsilon_\text{bas}$ comes from the regularized L{\"o}wdin orthonormalization transformation (\ref{eq:lowdindepfix}),  and is about removing completely a subspace from the basis set. Next we have the projector (\ref{eq:proj}) associated with an eigenvalue threshold $\epsilon_K$.  We can independently apply this to the exchange matrix $K$ through the similarity tranformation (\ref{eq:Kmatproj}) for Hartree Fock calculations, and to the orbital coefficients like in equation (\ref{eq:transformed-coefficients}) entering the MP2 and RPA energy expressions. We do not consider the possibility of using different values for $\epsilon_K$ for Hartree Fock, MP2 and RPA, other than completely bypassing the projector. Finally, when constructing an automatic fit set for an atom type, again a regularized L{\"o}wdin orthonormalization is used, based on the eigensystem of the overlap matrix of the fit functions now in the Coulomb metric, controlled by the eigenvalue threshold $\epsilon_\text{fit}$. This last threshold controls the number of fit functions, see Alg. \ref{algo:fit}. The same threshold is also employed for the pseudo inverse to obtain the PADF fit coefficients.
\section{\label{sec::comp}Computational Details}
All calculations have been performed with modified development versions of the ADF\cite{ADF2019} and BAND\cite{band22} modules of AMS2022.\cite{Ruger2022} All Psi4\cite{psi4} calculations have been performed using version 1.6.1.

\subsection{AMS calculations}
For all calculations using GTOs we used correlation consistent Dunning basis sets of double-$\zeta$ (DZ), triple-$\zeta$ (TZ), quadruple-$\zeta$ (QZ) and 5Z quality\cite{dunning89} from the basis set exchange library.\cite{Pritchard2019} For comparison with ADF, we used the triple-$\zeta$ plus double polarization (TZ2P) basis set.\cite{vanlenthe03}

In all BAND calculations we used the PADF scheme for the Hartree-Fock exchange operator while the Hartree potential fitting procedure is based upon a partitioning of the density in atomic reservoirs each of which is expanded in products of radial splines and spherical harmonics.\cite{franchini14} This procedure does not rely in any way on the PADF fit functions.
Throughout this work we performed tests varying the threshold $\epsilon_\text{fit}$ to control the size of the fit set. The thresholds used for particular calculations will be indicated in the next section. The same holds for the threshold for the canonical orthogonalization of the primary basis. If not indicated otherwise, we set the numerical quality to \emph{VeryGood}, which controls the accuracy of the numerical integration,\cite{franchini13} the quality of the ZLMfit,\cite{franchini14} as well as of the threshold controlling distance effects in HF, MP2 and RPA calculations.\cite{Forster2020} The same settings have been used in all ADF calculations. In order to make the basis functions more compact, in BAND the radial part of the basis functions is multiplied by a Fermi-Dirac (FD) function by default. We disabled this behavior in all calculations. In all RPA calculations for the S66 dataset we calculated the polarizability directly in imaginary frequency using \eqref{rpa_canonical} and modified Gauss--Legendre grids as described in ref.~\citen{Fauser2021} with 50 integration points. 

In all calculations for the L7\cite{Sedlak2013} and S30L\cite{Sure2015} datasets we used the imaginary time based algorithms for RPA and SOS-MP2 algorithm. In all SOS-MP2 calculations we used 12 imaginary time points which ensures $\mu$Hartree convergence of correlation energies of organic systems with large HOMO-LUMO gaps.\cite{Doser2009a,Maurer2014a} In all RPA calculations\cite{Forster2020b,Forster2022a} we used 24 imaginary frequency and imaginary time points each and and used PBE orbitals as input (RPA@PBE).

We calculated the interaction energies of the dimers in the CIM8 dataset\cite{Ni2021} at the RPA@PBE level of theory. If not indicated otherwise we used correlation consistent Dunning basis sets of DZ and TZ quality. We then extrapolated the final correlation energies to the complete basis set limit using the relation,\cite{Helgaker1997}
\begin{equation}
    \label{helgaker}
    E_\text{CBS} = E_{xZ} + \frac{E_{xZ} * x^3 - E_{(x-1)Z} * (x-1)^3}{x^3-(x-1)^3} \;, 
\end{equation}
where $x = 3$ for TZ, $x=4$ for QZ, and so on.
We set the numerical quality to \emph{Good} and set the threshold controlling distance effects for the RPA calculation to \emph{Normal}. Also here we used various settings for the quality of the fit set. For details we refer to the next section. For reasons discussed in the next section, if not indicated otherwise we set $\epsilon_K = 10^{-2}$ and $\epsilon_{\text{bas}}=5 \cdot 10^{-4}$ for calculations on the CIM8 dataset. Detailed input settings for all calculations can be found in the supporting information.

\subsection{Psi4 calculations}
We performed Psi4 calculations for the S66 database using Dunning double-$\zeta$ (DZ), triple-$\zeta$ (TZ), quadruple-$\zeta$ (QZ) to quintuple-$\zeta$ (5Z) basis set, to perform global DF-MP2 calculations (in the following simply referred to as DF-MP2).
We used default settings for all calculations, We used the default fit sets for each basis set, i.e. cc-pvxZ-RI\cite{Weigend2002a,Hattig2005,Hill2008} for the primary basis cc-pvxZ. 

\section{\label{sec::res}Results}
\label{sec:results}

In this section, we assess the accuracy of the algorithms described herein. We proceed as follows: In section~\ref{ssec::hf-pm} we first illustrate the effect of the threshold chosen for the regularized L{\"o}wdin orthogonaliazation (basis set reduction, $\epsilon_\text{bas}$) as well as for the HF projector method ($\epsilon_K$), on the exchange matrix, for a simple molecule. In the subsequent sections, we compare our results for molecules of increasing size, starting with the S66 database in section~\ref{ssec::s66} and moving on to the L7 and S30L databases which contain molecules with more than 200 atoms in section~\ref{ssec::large}. Finally, in section~\ref{ssec::CIM8} we showcase the capabilities of our PADF based algorithms by calculating the interaction energies of 7 large non-covalently bound complexes in the CIM8 set by Neese and coworkers with up to 910 atoms (4500 electrons).\cite{Ni2021}

\subsection{\label{ssec::hf-pm}Effect of the HF Projector Method}

\begin{figure}
    \centering
    \includegraphics[width=0.6\textwidth]{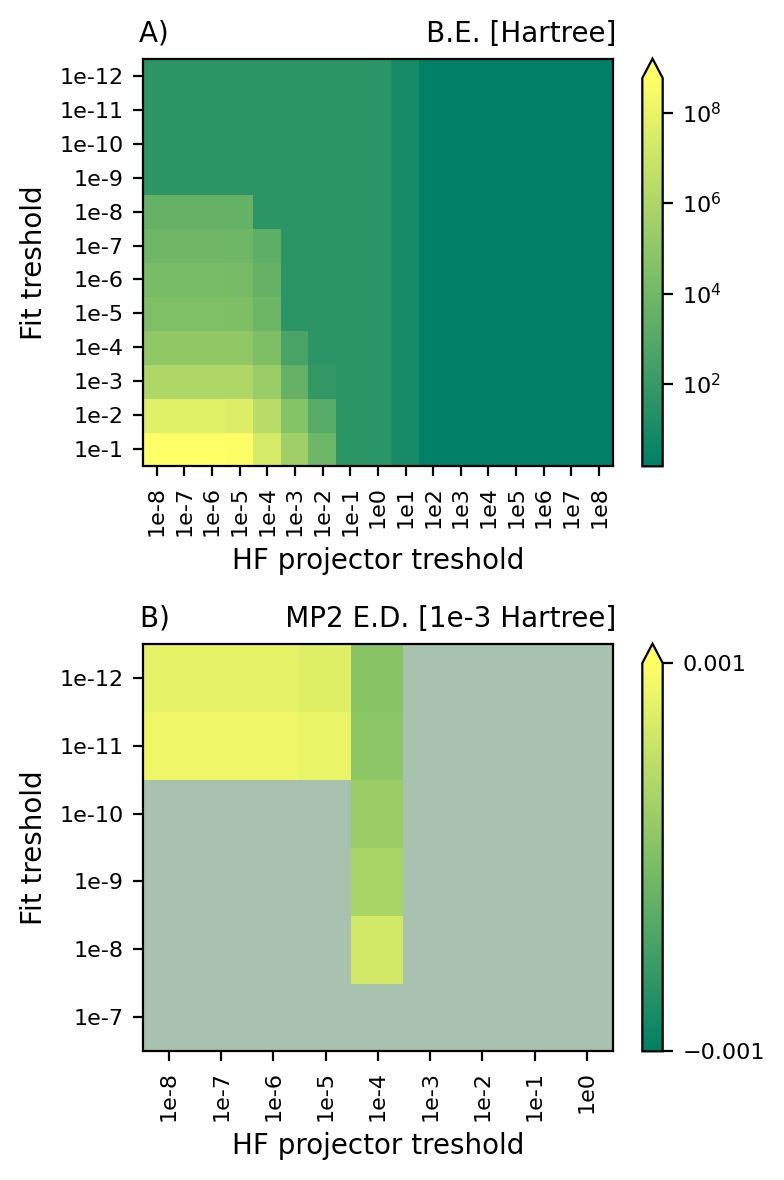}
    \caption{Comparison fit quality $\epsilon_\text{fit}$ with PM threshold $\epsilon_K$, in A figure we plot bonding energy of fluorobenzene, in B the deviation from its MP2 correlation energy computed through Psi4. Picture B contains a subset of the grid checked in A.}
    \label{fig:proj_grid}
\end{figure}
We start the discussion of our results by illustrating the effect of the HF projector method on the PADF-MP2 correlation energy of the fluorobenzene molecule for varying size of the fit set. In the heatmaps in Fig.~\ref{fig:proj_grid}A and Fig.~\ref{fig:proj_grid}B, we show the MP2 bonding energy and MP2 correlation energy of fluorobenzene for different thresholds for the HF PM $\epsilon_K$ and fit quality $\epsilon_\text{fit}$. 
In particular, in \ref{fig:proj_grid}A we report the bonding energy and highlight three interesting zones in the heatmap. On the right side, the PM threshold $\epsilon_K$ is large enough to completely remove the exchange energy thus reducing to the Hartree limit of HF. In the lower left corner where the quality of the fit set is poor, the PM threshold does only have a small effect and the PADF-MP2 bonding energy collapses to unphysically low values. The rest of the heatmap shows a more stable behavior, but still a broad range of bonding energies. To check the accuracy of the PADF-MP2 implementation in BAND, in Fig.~\ref{fig:proj_grid}B  we show values deviating less than 0.1 \% from the reference DF-MP2 correlation energy (Psi4), and we blur in grey the ones outside of the range. Thus we observe that increasing the threshold $\epsilon_{K}$, allows to use smaller fit sets while still maintaining a good precision in the result. 
 
\subsection{\label{ssec::s66}Deviation BAND-Psi4}

\begin{figure}
    \centering
    \includegraphics[width=0.6\textwidth]{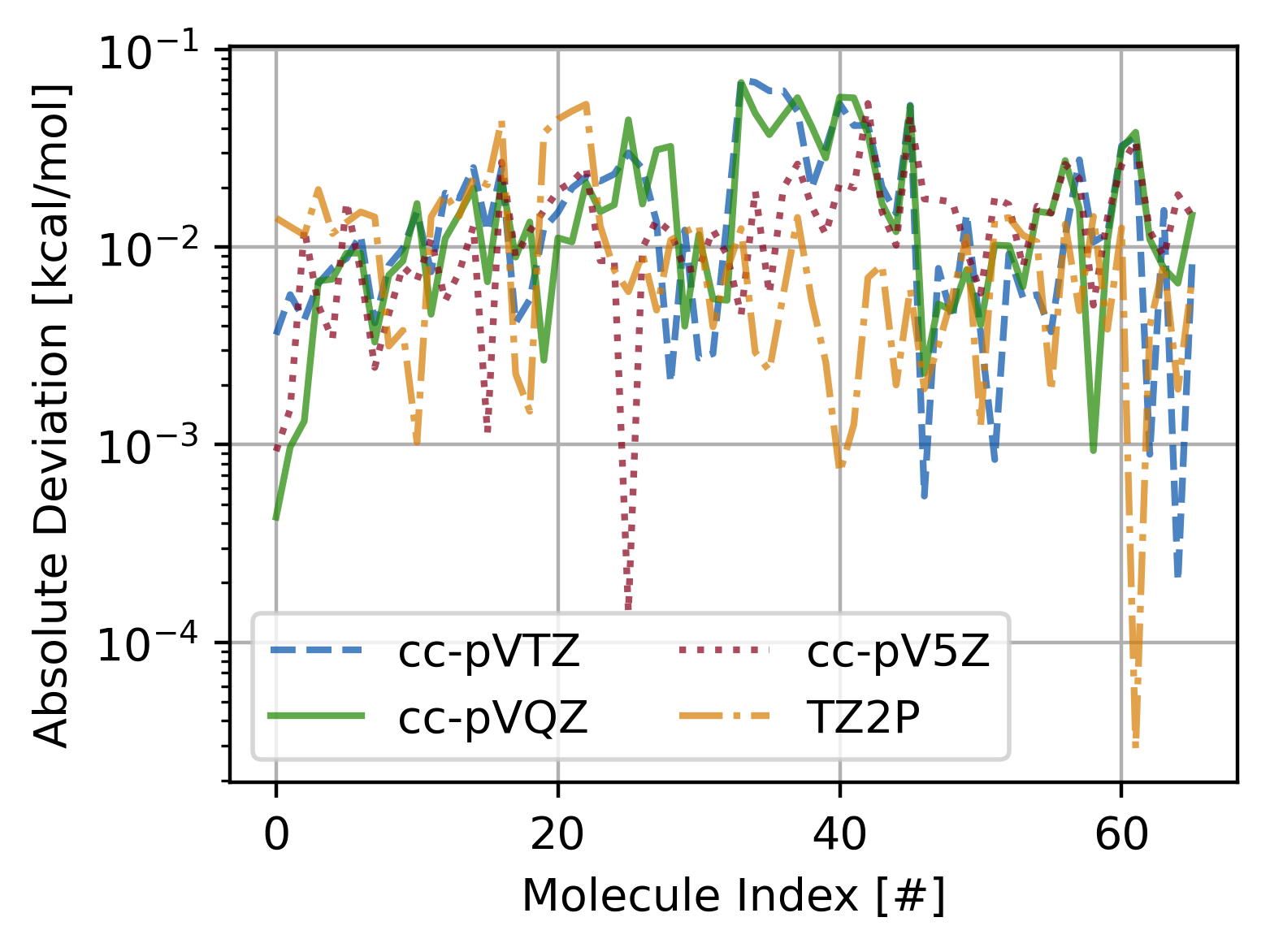}
    \caption{Absolute differences in non-covalent interaction energies between Psi4 and BAND and BAND and ADF for the S66 database using different basis sets. All values expressed in $\nicefrac{\text{kcal}}{\text{mol}}$}. 
    \label{fig:dev_int_psi4_band_hf_mp2_s66}
\end{figure}

 \begin{table}[ht]
     \centering
     \begin{tabular}{llccl}
          Codes & Basis Set & Max. & MAD & Worst molecule\\
          \hline \hline
           \multirow{3}{*}{$\Delta$MP2}& cc-pVTZ & 0.07 & 0.02 & Pentane-Pentane\\
           & cc-pVQZ  & 0.06 & 0.02 & Pentane-Pentane\\ 
           & cc-pV5Z  & 0.05 & 0.01 & Uracil-Cyclopentane \\ 
          $\Delta$ (ADF-BAND) & TZ2P         & 0.05 & 0.01 & AcOH-Uracil \\ 
         $\Delta$ RPA & cc-pVTZ & 0.14 & 0.05 & Uracil-Uracil ($\pi-\pi$) \\ 
     \end{tabular}
     \caption{Maximum deviations and Mean absolute deviations (MAD) between different PADF and DF implementations of MP2 and RPA for the non-covalent interaction energies in the S66 database for different basis sets. All values expressed in $\nicefrac{\text{kcal}}{\text{mol}}$.: First three rows: Deviation of DF-MP2 (Psi4) and PADF-MP2 (BAND), fourth row: Deviations of the PADF-MP2 implementations in ADF and BAND, last row: Deviations of PADF-RPA (BAND) to DF-RPA (TURBOMOLE) corrected by the frozen core error (see text for explanation).}
     \label{tab:maxerr}
 \end{table}
We now compare our PADF-MP2 and PADF-RPA results to DF-MP2 and DF-RPA reference values for the non-covalent interaction energies of the S66 database. We calculated the DF-MP2 reference values using Psi4. As reference values for PADF-RPA we use the TURBOMOLE results by Furche and coworkers\cite{Nguyen2020} who used the frozen core approximation. Comparing our DF-MP2 results obtained with Psi4 to the DF-MP2 results of TURBOMOLE, we found the error introduced by the frozen core approximation to be between 0.0 and 0.1 $\nicefrac{\text{kcal}}{\text{mol}}$ for relative energies. Therefore, to allow for a better comparison of our PADF-RPA to the DF-RPA results, we corrected the latter ones by the frozen core error for MP2, assuming the impact of the frozen core approximation to be the same for MP2 and RPA. We stress that this procedure might not remove the frozen core error completely and therefore the DF-RPA reference values have certainly higher errors bars than the DF-MP2 ones

In our calculations the relevant thresholds are the ones for the regularized L\"{o}wdin orthonormalization $\epsilon_\text{bas}$, the projector method, $\epsilon_K$ and for the size of the fit set, $\epsilon_\text{fit}$. For the MP2 calculations, we used $\epsilon_\text{bas} = 10^{-8}$ We set the threshold for the the projector method (PM) to \texttt{$\epsilon_{K} = 10^{-3}$}, and we used $\epsilon_\text{fit}=10^{-12}$ except for some of the $5\zeta$ calculations for which we used $10^{-10}$ instead (see supporting information for details). The $L$-e, was enabled only for $3\zeta$ calculations. This corresponds to an (unrealistically sized) fit set which is around 15 times larger than the primary basis.

The absolute deviations (AD) for all PADF-MP2 interaction energies in the S66 database with respect to DF-MP2 are shown in Fig.~\ref{fig:dev_int_psi4_band_hf_mp2_s66} for Dunning basis sets of $3\zeta$ to $5\zeta$ quality. In the same plot, we also show the deviations of the PADF-MP2 implementations in ADF and BAND using STOs. We see that in all cases, the deviations are smaller than 0.1  $\nicefrac{\text{kcal}}{\text{mol}}$, irrespective of the basis set. The small deviations between BAND and Psi4 should primarily be due to errors introduced by the PADF approximation. Given that the two implementations differ also in other aspects, we can, however, not exclude the possibility that differences in other technical parameters might play a role as well, for instance differences in the definitions of the numerical integration grids. Also incompleteness of the fit sets used in Psi4 might play a role. The deviations of BAND to ADF are mostly due to slightly different integration grids. In any case, for all practical purposes the agreement between the codes is excellent. Therefore, we did not investigate the precise origin of these small discrepancies further. The results of these calculations are summarized in table~\ref{tab:maxerr}.

In table~\ref{tab:maxerr}, we show that both maximum error and MAD of PADF-RPA relative to DF-RPA are about twice as large as for PADF-MP2. The reason for this might be that the fit errors are more pronounced due to the presence of the higher powers of $\mathbf{Z}$ (see \eqref{eq:Z}). There is also a slightly larger uncertainty in the reference values due to our approach to subtract to the frozen core errors in the reference calculations. we also assessed the effect of using a higher threshold of $\epsilon_{\text{bas}} = 10^{-5}$. This did however not change our results at all, demonstrating the numerical stability of our method.

\begin{figure}[hbt!]
    \centering
    \includegraphics[width=0.6\textwidth]{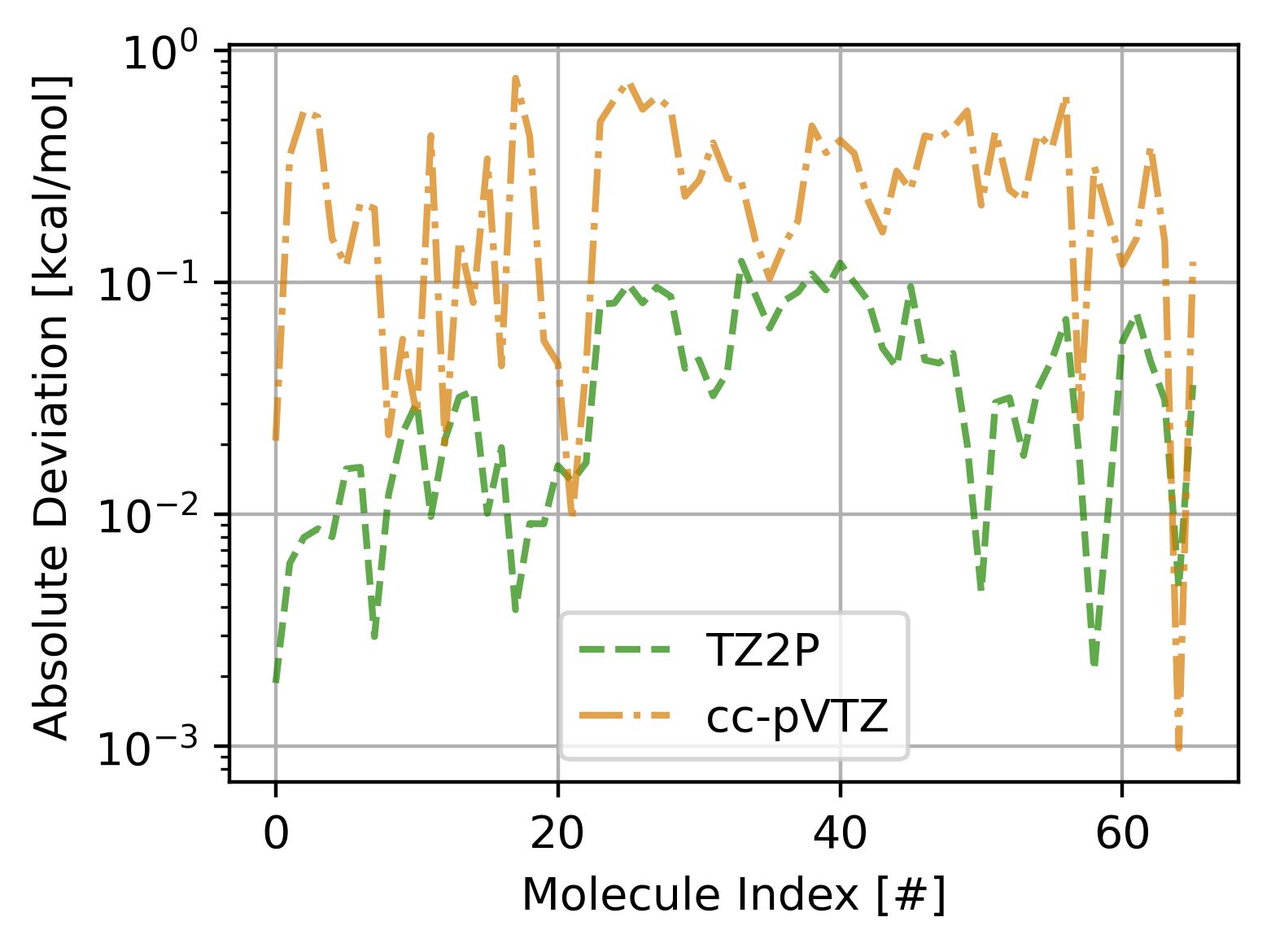}
    \caption{Differences in MP2 Interaction energies of the S66 dataset with and without $l$-e in the fit set. All values are in kcal/mol.}
    \label{fig:boostL}
\end{figure}
The $l$-e procedure is of key importance for both the GTO and STO triple zeta basis sets. This is illustrated in Fig.~\ref{fig:boostL} where we show the deviations of the PADF-MP2 interaction energies with and without $l$-e. For the cc-pVTZ basis set the $l$-e causes differences of the order of $\sim 1.0 \nicefrac{\text{kcal}}{\text{mol}}$ and of $\sim 0.1 \nicefrac{\text{kcal}}{\text{mol}}$ for TZ2P. The larger deviation in cc-pVTZ from using $l$-e can be traced back to the fact that an STO is added to a GTO basis set, thus adding functions with a different radial behaviour to the fit. For the TZ2P basis set the only difference comes from the higher angular momentum in the basis. The same comparison of PADF-RPA correlation energies gave similar results.

\subsection{Comparison of projector methods}

\begin{figure}[hbt!]
    \centering
    \includegraphics[width=0.6\textwidth]{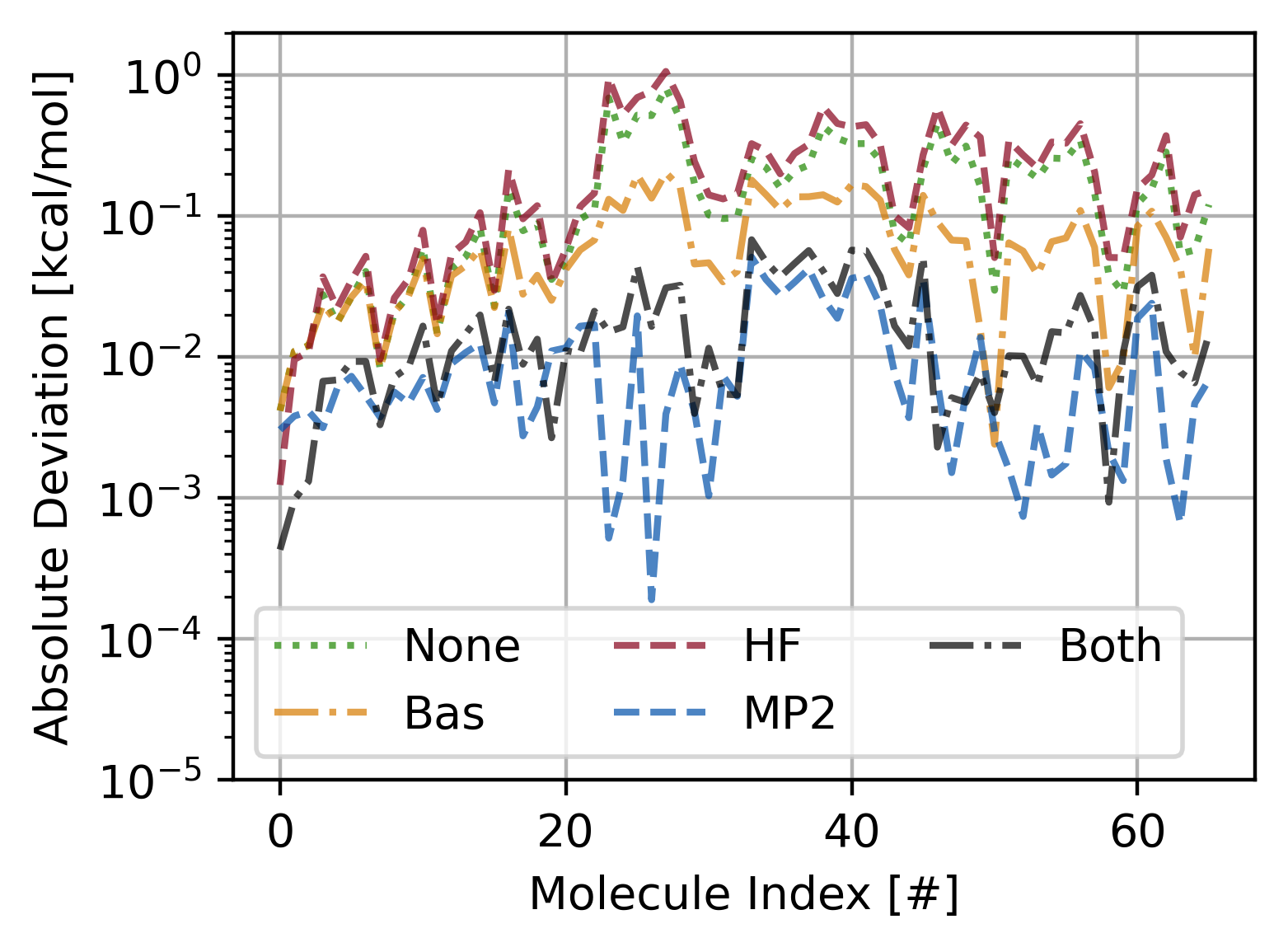}
     \centering
     \begin{tabular}{cccc}
          Label & $\epsilon_K$ (HF) & $\epsilon_K$ (MP2) & $\epsilon_\text{bas}$\\
          \hline \hline
         None & $\times$ & $\times$ & 1e-8 \\
         Bas  & $\times$ & $\times$ & 1e-4 \\
         HF   & 1e-3 & $\times$ & 1e-8\\
         MP2  & $\times$ & 1e-3  & 1e-8\\ 
         Both & 1e-3 & 1e-3 & 1e-8
         
     \end{tabular}

\caption{Absolute differences in non-covalent MP3 interaction energies between Psi4 and BAND for the S66 database using the cc-pVQZ basis set and yarying parameters for the different projector methods. Each line in the figure correspond to a different combination of the three parameters in the table. Symbol $\times$ means that the respective method has not been used. All values are in $\nicefrac{\text{kcal}}{\text{mol}}$.} 
    \label{fig:proj}
\end{figure}

In Fig.~\ref{fig:proj} we compare BAND results using several combinations of the above mentioned thresholds and show absolute deviation with respect to Psi4 for the S66 dataset.  
The names for the different parameters combinations are specified in the table at the bottom of the same figure. We compare results obtained without any projector(\texttt{None}), without PM-$K$ but using a tighter threshold for the regularized L\"{o}wdin orthonormalization in the basis set (\texttt{Bas}), using PM-$K$ only at the HF or at the MP2 stage, and finally using PM-$K$ in both cases (mentioned following the order of the table). 
The best results are obtained applying PM-$K$ to MP2 only, followed by applying it to both. This shows that the usage of PM-$K$ at the HF stage does not contribute much to the overall accuracy and in fact can even worsen it. This supports our hypothesis from section~\ref{sec::theo} since correlation energies are dependent on virtual orbitals which have more nodes and are more diffuse than the occupied ones. Hence, they are more difficult to represent in terms of an atomic centered basis and consequently their products more difficult to express in terms of fit functions. We emphasize however that we use here a very large fit set. When fewer fit functions are used, PM-$K$ also needs to be used at the SCF stage to prevent variational collapse. The calculations using a larger value of the L\"{o}wdin orthonormalization threshold lead to improvements with respect to the \texttt{None} and \texttt{HF} settings since unstable components of the basis are projected out from all terms of the Fock matrix. This is decisive for the accuracy of MP2 and RPA correlation energies as we have already seen in the comparisons above.

\subsection{\label{ssec::large}Interaction energies for L7 and S30L}
After having demonstrated the excellent agreement of our PADF-MP2/RPA results with DF-MP2/RPA also for large basis sets up to 5Z quality, we now discuss the accuracy of PADF-MP2 for the molecules in the L7 and S30L databases, for which Furche and coworkers have recently published DF-MP2 reference values.\cite{Nguyen2020} We focus on the direct contribution to the MP2 correlation energy only and calculate SOS-MP2 interaction energies using our imaginary time based PADF-SOS-MP2 implementation.\cite{Forster2020} To represent the imaginary time dependence we chose here 12 integration points in the interval $[0,\infty)$ which is sufficient to achieve $\mu$H precision in absolute correlation energies. The reference values by Furche and coworkers have been calculated with the frozen core approximation.\cite{Nguyen2020} We have however seen for the S66 database, that the maximum error of this approximation is of the order of 0.1 $\nicefrac{\text{kcal}}{\text{mol}}$ for relative energies and we do not expect this to change for larger molecules. Finally, Furche and coworkers only calculated correlation energies using Dunning basis sets while they used basis sets of the Kahlsruhe type to calculate their HF or KS energies.\cite{Nguyen2020} Therefore, we do not compare the non-covalent interaction energies for these sets but only the correlation contributions to them.

\subsubsection{L7}

\begin{table}[hbt!]
    \centering
    \begin{tabular}{l
    S[table-format=3.2]%
    S[table-format=3.2]%
    S[table-format=3.2]%
    S[table-format=3.2]%
    S[table-format=3.2]%
    S[table-format=3.2]%
    S[table-format=3.2]%
    S[table-format=3.2]%
    S[table-format=3.2]%
    }
    \toprule
  {$\epsilon_K$}  & {$l_\text{max}$} & \multicolumn{4}{c}{Non-covalent interaction energy} & \multicolumn{4}{c}{$\Delta_{\text{TURBOMOLE}}$} \\
{$1e-3$} & {6} \\ \cline{1-2}
{System}   &  {Ref.} &   {1e-6} &  {1e-8} & {1e-10} & {1e-12} &  {1e-6} &  {1e-8} & {1e-10} & {1e-12} \\
\midrule
c2c2PD                         &   -33.72 &   -34.92 &   -34.46 &   -34.20 &   -34.22 &    -1.20 &    -0.74 &    -0.48 &    -0.50 \\
c3a                            &   -22.07 &   -22.96 &   -22.65 &   -22.37 &   -22.53 &    -0.89 &    -0.58 &    -0.30 &    -0.46 \\
c3gc                           &   -37.81 &   -39.37 &   -38.81 &   -38.38 &   -38.85 &    -1.56 &    -1.00 &    -0.57 &    -1.04 \\
cbh                            &   -11.49 &   -13.19 &   -12.61 &   -12.38 &   -12.26 &    -1.70 &    -1.12 &    -0.89 &    -0.77 \\
gcgc                           &   -18.44 &   -19.64 &   -19.14 &   -18.92 &   -22.37 &    -1.20 &    -0.70 &    -0.48 &    -3.93 \\
ggg                            &    -7.96 &    -8.32 &    -8.21 &    -8.13 &    -8.59 &    -0.36 &    -0.25 &    -0.17 &    -0.63 \\
phe                            &    -5.15 &    -5.87 &    -5.63 &    -5.51 & {--}     &    -0.72 &    -0.48 &    -0.36 &     {--} \\
\rule{0pt}{4ex}  {$\epsilon_K$}  & {$l_\text{max}$} & \multicolumn{4}{c}{Non-covalent interaction energy} & \multicolumn{4}{c}{$\Delta_\text{TURBOMOLE}$} \\
{$1e-3$} & {8} \\ \cline{1-2}
{System}   &  {Ref.} &   {1e-6} &  {1e-8} & {1e-10} & {1e-12} &  {1e-6} &  {1e-8} & {1e-10} & {1e-12} \\
\midrule
c2c2PD                         &   -33.72 &   -34.82 &   -34.23 &   -33.75 &          &    -1.10 &    -0.51 &    -0.03 \\
c3a                            &   -22.07 &   -22.90 &   -22.50 &   -28.58 &          &    -0.83 &    -0.43 &    -6.51 \\
c3gc                           &   -37.81 &   -39.26 &   -38.45 &   -38.06 &          &    -1.45 &    -0.64 &    -0.25 \\
cbh                            &   -11.49 &   -12.96 &   -12.25 &   -11.94 &          &    -1.47 &    -0.76 &    -0.45 \\
gcgc                           &   -18.44 &   -19.53 &   -18.97 &   -29.75 &          &    -1.09 &    -0.53 &   -11.31 \\
ggg                            &    -7.96 &    -8.29 &    -8.15 &    -8.37 &          &    -0.33 &    -0.19 &    -0.41 \\
phe                            &    -5.15 &    -5.82 &    -5.53 &    -6.02 &          &    -0.67 &    -0.38 &    -0.87 \\
\rule{0pt}{4ex}  {$\epsilon_K$}  & {$l_\text{max}$} & \multicolumn{4}{c}{Non-covalent interaction energy} & \multicolumn{4}{c}{$\Delta_\text{TURBOMOLE}$} \\
{$5e-3$} & {6} \\ \cline{1-2}
{System}   &  {Ref.} &   {1e-6} &  {1e-8} & {1e-10} & {1e-12} &  {1e-6} &  {1e-8} & {1e-10} & {1e-12} \\
\midrule
c2c2PD                         &   -33.72 &   -34.02 &   -33.80 &   -33.69 &   -33.69 &    -0.30 &    -0.08 &     0.03 &     0.03 \\
c3a                            &   -22.07 &   -22.15 &   -22.04 &   -21.96 &   -21.97 &    -0.08 &     0.03 &     0.11 &     0.10 \\
c3gc                           &   -37.81 &   -37.95 &   -37.76 &   -37.63 &   -37.67 &    -0.14 &     0.05 &     0.18 &     0.14 \\
cbh                            &   -11.49 &   -12.02 &   -11.85 &   -11.78 &   -11.71 &    -0.53 &    -0.36 &    -0.29 &    -0.22 \\
gcgc                           &   -18.44 &   -18.46 &   -18.35 &   -18.29 &   -18.60 &    -0.02 &     0.09 &     0.15 &    -0.16 \\
ggg                            &    -7.96 &    -7.95 &    -7.92 &    -7.91 &    -7.93 &     0.01 &     0.04 &     0.05 &     0.03 \\
phe                            &    -5.15 &    -5.31 &    -5.22 &    -5.19 &    -8.18 &    -0.16 &    -0.07 &    -0.04 &    -3.03 \\
\bottomrule
    \end{tabular}
    \caption{Comparison of SOS-MP2 contributions to the non-covalent interaction energies in the L7 database to the ones from Furche and coworkers\cite{Nguyen2020} for different numerical settings using the cc-pvTZ basis set. All values are in $\nicefrac{\text{kcal}}{\text{mol}}$. The fit sets have been generated from the basis products using the threshold $\epsilon_{\text{fit}}$}
    \label{tab::l7_sosmp2}
\end{table}

For the fit sets generated from the products of basis functions, the results of this comparison for the L7 database can be found in table~\ref{tab::l7_sosmp2}. We first focus on the upper table. The results here have been obtained using a value of $\epsilon_K = 10^{-3}$ as threshold for PM-$K$ (the same value as for S66) and without the $l$-e method. The results for the different values of $\epsilon_k$ controlling the size of the fit set ranging from $10^{-6}$ to $10^{-12}$ in table~\ref{tab::l7_sosmp2} show a slow convergence of the relative SOS-MP2 correlation energies to the DF-SOS-MP2 reference values. However, even with the already very large fit set corresponding to $\epsilon_k=10^{-10}$ (the number of fit functions is around 10 times larger than the number of primary basis functions), the maximum deviation is still 0.89 $\nicefrac{\text{kcal}}{\text{mol}}$ and only for the smallest of the systems (GGG) in L7, the deviation to the TURBOMOLE results reaches an acceptable value of 0.17 $\nicefrac{\text{kcal}}{\text{mol}}$. Moreover, the results for $\epsilon_k=10^{-12}$ start to worsen, which can be related to the occurrence of linear dependencies in the fit set. For the Phe system in L7 we even obtained a completely unreasonable value for one of the correlation energies.

Turning to the second table in table~\ref{tab::l7_sosmp2}, we find that the $l$-e reduces the error with respect to DF-SOS-MP2 compared to the value obtained with the same thresholds. However, we already observe numerical instabilities for fit sets corresponding to $\epsilon_k=10^{-10}$. Notice however, that the total size of this fit set is already larger than the one corresponding to $\epsilon_k=10^{-12}$ without the $l$-e. Therefore, also the $l$-e method does not resolve the issues of PADF for the molecules in L7. 

We now turn to the third table in table~\ref{tab::l7_sosmp2} which shows results obtained without the $L$-e method but with a larger threshold of the HF projector method, $\epsilon_K=5\cdot 10^{-3}$, instead of $\epsilon_K=10^{-3}$ which has been used in the two previous tables. This changes the PADF-SOS-MP2 correlation energies drastically and bring them into much better agreement with the TURBOMOLE results. Already for a moderate size of the fit set corresponding to $\epsilon_k=10^{-6}$, the maximum deviation is reduced to 0.53 $\nicefrac{\text{kcal}}{\text{mol}}$ and for a value of $\epsilon_k=10^{-8}$, for 6 out of 7 systems the agreement with the reference values is better than 0.1 $\nicefrac{\text{kcal}}{\text{mol}}$. Notice again, that 0.1 $\nicefrac{\text{kcal}}{\text{mol}}$ is of the order of uncertainty in the reference values due to the frozen core approximation. The CBH complex is the only system for which the deviation to the reference is still relatively large. Only with $\epsilon_k=10^{-12}$, the deviation to the TURBOMOLE results reduces to an acceptable value of $0.22 \nicefrac{kcal}{mol}$. However, first this fit set is very large and therefore not very useful in applications to large molecules, and second, we can still observe quite large deviations to TURBOMOLE data for other systems, most drastically for PHE.

\begin{table}[hbt!]
    \centering
    \begin{tabular}{l
    S[table-format=3.2]%
    S[table-format=3.2]%
    S[table-format=3.2]%
    S[table-format=3.2]%
    S[table-format=3.2]%
    S[table-format=3.2]%
    S[table-format=3.2]%
    S[table-format=3.2]%
    }
    \toprule
  {$\epsilon_K=5\times 10^{-3}$} & \multicolumn{4}{c}{SOS-MP2} & \multicolumn{4}{c}{RPA} \\
{System} &     {$E^{\text{SOS-MP2}}_{\text{Ref.}}$} &  {normal} &     {good} &  {vg} &{$E^{\text{RPA}}_{\text{Ref.}}$} & {normal} & {good} & {vg} \\
\midrule
c2c2PD     &   -33.72   &    -2.09 &    -0.41 &     0.09 &   -30.29 &  -2.17 & -0.32 &  0.30 \\
c3a        &   -22.07   &    -1.32 &    -0.20 &     0.14 &   -21.21 &  -1.43 & -0.12 &  0.32 \\
c3gc       &   -37.81   &    -2.16 &    -0.30 &     0.27 &   -39.89 &  -0.19 &  2.02 &  2.76 \\
cbh        &   -11.49   &    -0.83 &    -0.10 &    -0.03 &   -16.84 &  -0.94 & -0.03 &  0.10 \\
gcgc       &   -18.44   &    -0.72 &    -0.05 &     0.21 &   -22.33 &  -0.89 &  0.04 &  0.37 \\
ggg        &    -7.96   &    -0.28 &    -0.03 &     0.07 &    -9.00 &  -0.32 &  0.02 &  0.15 \\
phe        &    -5.15   &    -0.51 &    -0.04 &     0.02 &    -8.53 &  -0.60 & -0.01 &  0.10 \\
\rule{0pt}{4ex}    {$\epsilon_K=10^{-2}$} & \multicolumn{4}{c}{SOS-MP2} & \multicolumn{4}{c}{RPA} \\
{System} &     {$E^{\text{SOS-MP2}}_{\text{Ref.}}$} &  {normal} &   {good} & {vg} & {$E^{\text{RPA}}_{\text{Ref.}}$}& {normal} & {good} & {vg} \\
\midrule
c2c2PD &   -33.72  & -1.31 &    -0.01 &     0.09 &   -30.29 & -1.24 &     0.22 &    0.71 \\
c3a    &   -22.07  & -0.84 &     0.06 &     0.14 &   -21.21 & -0.76 &     0.28 &    0.62 \\
c3gc   &   -37.81  & -1.37 &     0.11 &     0.27 &   -39.89 &  0.90 &     2.68 &    3.25 \\
cbh    &   -11.49  & -0.73 &    -0.07 &    -0.03 &   -16.84 & -0.73 &     0.09 &    0.17 \\
gcgc   &   -18.44  & -0.57 &     0.07 &     0.21 &   -22.33 & -0.66 &     0.28 &    0.55 \\
ggg    &    -7.96  & -0.17 &     0.02 &     0.07 &    -9.00 & -0.15 &     0.12 &    0.22 \\
phe    &    -5.15  & -0.43 &    -0.05 &     0.02 &    -8.53 & -0.43 &     0.04 &    0.11 \\
\bottomrule
    \end{tabular}
    \caption{Deviations $\Delta_\text{TURBOMOLE}$ of SOS-MP2 (left) and RPA (right) contributions to the non-covalent interaction energies in the L7 database to the ones from Furche and coworkers\cite{Nguyen2020} for different numerical settings using the cc-pvTZ basis set. All values are in $\nicefrac{\text{kcal}}{\text{mol}}$. Standard STO type fit sets of varying size, randing from \emph{Basic} to \emph{VeryGood} (vg) quality have been used.}
    \label{tab::l7_sosmp2_2}
\end{table}

\paragraph{STO type fit sets}
Out of all systems in L7, CBH contains the largest number of atoms, but it is not the system with most electrons. Therefore, it can be considered as the most spatially extended system. As we have already discussed, such systems are expected to be most problematic for PADF based methods since many products of diffuse functions will occur. Since GTOs decay much faster than STOs from their atomic centres, it is natural to ask whether GTOs are the best choice to fit such products of diffuse atomic orbitals. Therefore, we investigate the accuracy which can be achieved by fitting the products of GTOs with STO type functions in table~\ref{tab::l7_sosmp2_2}. In the second table table of table~\ref{tab::l7_sosmp2_2} we show some results for $\epsilon_K=10^{-2}$. Additionally, we have reduced the integration quality to \emph{Good} and the quality of the threshold controlling distance effects in the SOS-MP2 method to \emph{Normal}. Both settings together drastically speed up the PADF-SOS-MP2 calculations. 

Using the\emph{Normal} fit set, relatively large errors are obtained. The results using the \emph{Good} and \emph{VeryGood} fit sets are in relatively good agreement with TURBOMOLE. Especially for the CBH complex, the deviation is much smaller than for the GTO type fit sets. Increasing $\epsilon_K$ to $10^{-1}$ improves agreement with the TURBOMOLE results further and using the \emph{Good} fit set results in perfect agreement with the reference values (given their uncertainties due to the frozen core approximation) are obtained.

Independently of the value of $\epsilon_K$, the errors of the relative RPA correlation energies are of the same order of magnitude as for SOS-MP2. However, as already observed for the S66 benchmark, the errors tend to be slightly larger. We also observe that relative energies tend to be less negative, indicating that smaller values of $\epsilon_K$ than for PADF-SOS-MP2 are beneficial for PADF-RPA results. Also here, using the \emph{Normal} fit set, errors hardly exceed 1 $\nicefrac{\text{kcal}}{\text{mol}}$. The system c3gc (A GC base pair absorbed on a Circumcoronene molecule) is particularly problematic with errors of the order of 3 $\nicefrac{\text{kcal}}{\text{mol}}$ for the \emph{VeryGood} fit set. We show in the supporting information that lowering of $\epsilon_K$ leads to much better agreement with experiment. This shows that the optimal settings for the projector methods are system specific and further research is needed to understand these patterns.

\paragraph{PADF-SOS-MP2 results using the CC-pVQZ basis set}

\begin{table}[hbt!]
    \centering
    \begin{tabular}{l
    S[table-format=3.2]%
    S[table-format=3.2]%
    S[table-format=3.2]%
    S[table-format=3.2]%
    S[table-format=3.2]%
    S[table-format=3.2]%
    S[table-format=3.2]%
    }
    \toprule
& & \multicolumn{3}{c}{Non-covalent interaction energy} & \multicolumn{3}{c}{$\Delta_\text{TURBOMOLE}$} \\
{System}   &  {Ref.} &   {1e-6} &  {1e-8} & vg & {1e-6} &  {1e-8} & vg \\
\midrule
c2c2PD                        &   -35.47&   -35.63&        & -36.66&  -0.16&       &  -1.19 \\
c3a                           &   -23.43&   -23.54&        & -24.15&  -0.11&       &  -0.72 \\
c3gc                          &   -40.33&   -40.58&        & -41.50&  -0.25&       &  -1.17 \\
cbh                           &   -12.27&   -12.64& -12.43 & -12.74&  -0.37& -0.16 &  -0.47 \\
gcgc                          &   -20.03&   -20.22&        & -20.53&  -0.19&       &  -0.50 \\
ggg                           &    -8.48&    -8.53&        &  -8.67&  -0.05&       &  -0.19 \\
phe                           &    -6.40&    -6.62&  -6.53 &  -6.74&  -0.22& -0.13 &  -0.34 \\
\bottomrule
    \end{tabular}
    \caption{Comparison of SOS-MP2 contributions to the non-covalent interaction energies in the L7 database to the ones from Furche and coworkers\cite{Nguyen2020} for different numerical settings using the CC-pvQZ basis set. All values are in $\nicefrac{\text{kcal}}{\text{mol}}$. All values are obtained using $\epsilon_K = 5\cdot 10^{-3}$.}
    \label{tab::l7_sosmp2_qz}
\end{table}

Lastly, we examine the quality of the PADF approximation at the QZ level. As for the S66 database, the results shown in table~\ref{tab::l7_sosmp2_qz} demonstrate that the quality of the interaction energies is not deteriorated compared to the TZ level. Already with the rather moderate threshold of $\epsilon_k = 10^{-6}$, the maximum deviation is -0.37 $\nicefrac{\text{kcal}}{\text{mol}}$ for CBH. This value is reduced to -0.16 $\nicefrac{\text{kcal}}{\text{mol}}$ when the threshold is decreased to $\epsilon_k = 10^{-8}$. Only using the \emph{VeryGood} Slater type fit set leads to worse results than at the TZ level.This is due to the fact that the \emph{VeryGood} fit set only contains functions with $l \leq 6$, with the number of $l=6$ functions being rather small, while the product basis contains functions with $l \leq 8$, with a large number of functions with $l \geq 6$.

\subsubsection{S30L}

\begin{figure}
    \centering
    \includegraphics[width=0.6\textwidth]{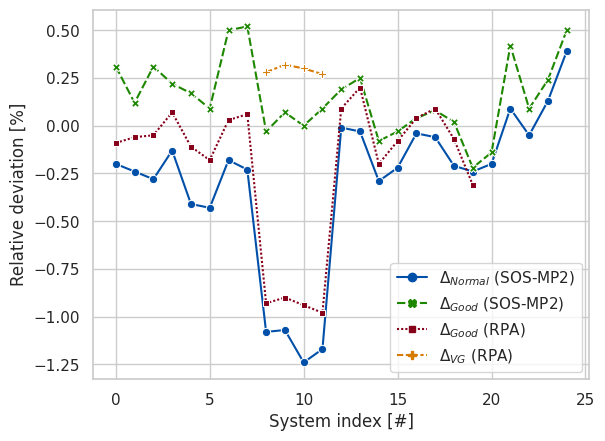}
    \caption{Deviations of PADF-SOS-MP2 and PADF-RPA with different STO type fit sets with respect to the TURBOMOLE reference values using global DF. All calculations have been performed using the cc-pvTZ basis set and the settings from table~\ref{tab::l7_sosmp2_2}. All deviations are in kcal/mol.}
    \label{fig::s30l}
\end{figure}

To further assess the quality of PADF-SOS-MP2, we also calculate the SOS-MP2 and RPA contributions to the interaction energies in the S30L set, which contains 30 non-covalently bound complexes with up to 205 atoms, out of which 8 are charged.\bibnote{We have omitted the systems 15 and 16 which contain Iodine atoms and for which a comparison to TURBOMOLE would be difficult due to the use of pseudopotentials in ref.~\cite{Nguyen2020}. We also do not show results for 3 of the charged systems due to convergence difficulties.} The deviations to the respective TURBOMOLE results are shown in figure~\ref{fig::s30l} and confirm the observations for the L7 database in table~\ref{tab::l7_sosmp2_2}. When the \emph{Good} fit set is used, the SOS-MP2 and RPA errors never exceed 0.5 and 1.0 kcal/mol, respectively. These results also demonstrate that the errors in relative energies do not seem to depend much on the sizes of the complexes any more after a certain system size is reached. This is in fact the expected results, since AOs centered on two atoms very far apart form each other will not overlap and therefore not contribute to the PADF errors.

\subsection{\label{ssec::CIM8}RPA Interaction energies of large complexes}

After having demonstrated the relatively good accuracy of the PADF approximation also for the correlation energies of large molecules, we now calculate the interaction energies in the CIM (Cluster-in-molecule)8 set by Neese and coworkers\cite{Ni2021} at the RPA@PBE level of theory. It comprises 8 large non-covalently bound complexes ranging in size from 200 to 1027 atoms, with interactions dominated by either $\sigma$-$\sigma$ dispersion or hydrogen bonding.\cite{Ni2021} 

Given the size of these systems, it is clear that these interaction energies can only be calculated if certain approximations are introduced. Neese and coworkers used an approach in which they decomposed the complex into smaller clusters and calculated the correlation energies using a cluster expansion of the general form
\begin{equation}
    E^\text{corr} = \sum_{I}E^\text{corr}_I + \sum_{I < J} E^\text{corr}_{IJ} \;,
\end{equation}
where $I$ denotes a subset of localized occupied and virtual molecular orbitals.\cite{Ni2021} They proposed to evaluate the correlation energy by two RI-MP2 calculations using the (aug-)cc-pvDZ and the (aug-)cc-pVTZ basis sets and by a DLPNO-CCSD(T) correction using the smaller basis. They used the augmented basis sets only for the two smallest systems while for the other systems they used the non-augmented basis sets.

Basis set incompleteness errors aside, there are major sources of inaccuracies with this approach, which could potentially lead to large errors. First, both the CIM and the DLPNO approximations can introduce errors in relative energies of several $\nicefrac{\text{kcal}}{\text{mol}}$ for non-covalent interactions, even when "tight" truncation thresholds for the DLPNO settings are used.\cite{Brandenburg2018, Carter-Fenk2019} Second, the chosen extrapolation scheme assumes that the CC basis set error can be faithfully estimated at the MP2 level. This is however not necessarily the case since the major basis set errors in CC calculations arise from the direct MP2 contribution,\cite{Petersson1981, Irmler2019, Irmler2019a} a observation termed as interference effect by Petersson et al.\cite{Petersson1988}  Especially since MP2 correlation energies will be rather inaccurate for the large molecules in CIM8, the strategy to estimate the basis set incompleteness error at the MP2 level will be error-prone. Finally, we mention the recently observed disagreement of well converged CCSD(T) interaction energies with quantum diffusion Monte Carlo methods for large non-covalently bound complexes.\cite{Al-Hamdani2021} For all these reasons, the values by Neese and coworkers are certainly not of quantitative accuracy. Despite all this, they are the most accurate reference values which are available for these large systems and certainly serve as useful frame of reference for our RPA@PBE calculations.

We have calculated all RPA@PBE interaction energies using \eqref{helgaker} using cc-pvDZ and cc-pVTZ basis sets with and without counterpoise corrections calculations. We have extrapolated the RPA correlation energies only, while the TZ results has been used for the remaining components of the interaction energies. We have verified the accuracy of this strategy by comparison to results using a (T,Q) extrapolation with the cc-pVTZ and cc-pVQZ basis sets. As shown in table S1 in the supporting information, the results of both extrapolation schemes differ by about 6.5 kcal/mol. This also indicates that for systems as large as the ones in CIM8 for which QZ calculations are out of reach, basis set incompleteness errors are typically much larger than the errors introduced by the PADF approximation. 

\begin{table}[hbt!]
    \centering
    \begin{tabular}{l
    S[table-format=5.2]%
    S[table-format=5.2]%
    S[table-format=3.2]%
    S[table-format=3.2]%
    S[table-format=3.2]%
    S[table-format=3.2]%
    S[table-format=3.2]}
    \toprule
    & & &
    \multicolumn{2}{c}{$E_{RPA}$ (D,T) (\% cp)}
    & & 
    \multicolumn{2}{c}{{$\Delta E$ (\% cp)}} \\ \cline{4-5} \cline{7-8}
    System 
    & {$N_{\text{atom}}$}
    & {$N_{\text{bas}}$}
    & {100 \% }
    & {0 \% }
    & {E(ref.) } 
    & {100 \%} 
    & {0 \%} \\ 
    \midrule
    1 &  200  & 5360  & -58.22	& -67.80  & -70.11 & 11.89 & 2.31 \\
    2 &  296  & 6576  & -55.76	& -62.79  & -63.61 & 7.85 & 0.82 \\
    4 &  328  & 9072  & -31.01	& -34.67  & -36.55 & 5.54 & 1.88 \\
    3 &  381  & 10806 & -14.09	& -24.90  & -17.83 & 3.74 & -7.07 \\
    5 &  552  & 12080 & -34.31	& -48.67  & -40.13 & 5.82 & -8.54 \\
    6 &  750  & 17316 & -58.97	& -86.81  & -78.80 & 19.83 & -8.01 \\
    7 &  910  & 21932 & -336.83	& -414.75 & -416.08 & 79.25 & 1.33 \\
    8 & 1027  & 22778 & -25.58	& -41.27  & -35.70 & 10.12 & -5.57 \\
    \cline{7-8}
    MAD:& & & & & & 18.00 & 4.44 \\
\bottomrule 
    \end{tabular}
    \caption{Interaction energies for eight large-non-covalently bound complexes. The number of basis functions refers to the full complex using cc-pVTZ. The ref. energy in the second last column has been taken from Neese and coworkers\cite{Ni2021} and has been calculated at the CIM-DLPNO-CCSD(T)|CIM-RI-MP2(D,T)Z level of theory.\cite{Ni2021} All interaction energies are in $\nicefrac{\text{kcal}}{\text{mol}}$ and have been extrapolated using \eqref{helgaker} with cc-pVDZ and cc-pVTZ. The counterpoise (cp) correction used is indicated by a percentage (0\% is no cp).}
    \label{tab:CIM8}
\end{table}

The results of our RPA@PBE calculations are shown in table~\ref{tab:CIM8}. The counterpoise corrected results are the most accurate RPA@PBE interaction energies we can calculate for these large systems. Given the good accuracy of RPA@PBE for large non-covalently bound complexes\cite{Nguyen2020}, they might serve as reference values for more approximate methods. However, they can not be compared directly to the values by Neese and coworkers since they did not correct for basis set superposition errors. For this purpose, we also calculated interaction energies which are not counterpoise corrected. 

Overall, reasonable agreement of our RPA@PBE results with the reference values by Neese and coworkers\cite{Ni2021} is observed.  With a MAD of 4.44 $\nicefrac{\text{kcal}}{\text{mol}}$, the deviations are of the same order of magnitude than popular (dispersion corrected) density functionals\cite{Wu2021, Ni2021, Bremond2022} (e.g. $\omega$B97-X-D with a MAD of 5.06 $\nicefrac{\text{kcal}}{\text{mol}}$ or B3LYP-D4 with a MAD of 4.81 $\nicefrac{\text{kcal}}{\text{mol}}$) and smaller than the ones for other \emph{ab initio} methods like SCS-MP2.\cite{Ni2021} We emphasize again that even though the agreement to the CIM-DLPNO-CCSD(T) reference is satisfactory, these non-counterpoise corrected interaction energies come with large basis set superposition errors and are therefore likely to be incorrect (see also table S2 in the supporting information).


 \section{\label{sec::sum}Conclusion}
By comparison to DF-MP2 and DF-RPA, we demonstrated the accuracy of PADF-MP2 and PADF-RPA for the S66, L7, and S30L sets of non-covalently bound complexes ranging from 6 to more than 200 atoms in size. Especially for the small to medium molecules in the S66 database, PADF comes with negligible loss of accuracy compared to global density fitting. 

The main advantage of PADF over global DF is that is leads to very fast algorithms for RPA and SOS-MP2. We have shown that the PADF approach is suitable to calculate the interaction energies of large molecules. In particular, we calculated the PADF-RPA@PBE interaction energies of eight large non-covalently bound complexes at the cc-pVTZ level with more than 1000 atoms and more than 20000 AOs in size on a single compute node. 

The choice of fit set is decisive for precise correlation energies with PADF. We tested two different types of fit sets. In a first variant, the fit set is generated directly from products of basis functions.\cite{Ihrig2015} In a second variant, fit sets consisting of even tempered series of STOs are used.\cite{Forster2020} Despite being much smaller, we found the second kind of fit set  to be suitable to express products of GTOs. This might be due to the slow decay of radial part of the STOs making them more suitable to fit delocalized products of AOs.

To improve the precision of the PADF approach further, we introduced a projector which acts directly on the Fock matrix and removes the attractive component of the excact exchange. most importantly, we also use this method to project out subspaces of AOs from the orbital coefficients matrix which can only be represented poorly by the fit set. Especially when smaller fit sets are used, we showed the PM-$K$ to be of key importance for accurate interaction energies.

While the PM reduces the correlation energy errors arising from PADF it cannot completely eliminate them. This is especially true for large molecules, for which a compromise between accuracy and computational efficiency is required. For very large systems like the molecules in the CIM8 dataset, it becomes mandatory to use smaller fit sets which might introduce errors in interaction energies which can exceed 1 \nicefrac{kcal}{mol}. This is however also true for many approximations to high-level methods for the calculation of correlation energies, for instance CC methods based on localized orbitals using the DLPNO\cite{Brandenburg2018, Carter-Fenk2019} or LNO approximations.\cite{Al-Hamdani2021} At the moment, the HF projector method is not used in a system specific way. A computationally efficient way to do so could be to check the definiteness of the Hartree-exchange matrix at runtime and to use this information to adjust $\epsilon_K$ during the SCF.

In practice, the errors stemming from the PADF approximation will often be of only minor relevance. Especially MP2 is typically not used by itself  but rather in double hybrids functionals which typically use a fraction of around 30-60 \% of MP2 correlation energy,\cite{Santra2019a, grimme06} scaling the error by the same amount. Therefore, the already small PADF errors will be negligible for those functionals. Furthermore, when small fit sets are used, the fit set incompleteness error always leads to too low interaction energies while basis set incompleteness errors lead to too high interaction energies. Especially for large molecules with several hundreds of atoms for which QZ calculations are not feasible, the fit set incompleteness errors will be much smaller than basis set incompleteness errors. 

On a more general note, it has recently been recognized that methods which agree well with each other for small and medium molecules might give very different results for larger systems.\cite{Al-Hamdani2021} Also, approximate (dispersion corrected) GGAs or hybrid functionals which work well for smaller systems are much more error-prone for large molecules.\cite{Wu2021,Bremond2022} In order to understand the reasons for this discrepancy, it is mandatory to push the boundaries of first-principle methods to much larger systems. At the moment, this comes at the price of numerical errors and further research is needed to develop techniques to mitigate these errors further.




\begin{acknowledgement}
We thank Erik van Lenthe for fruitful discussions and the idea behind the projector methods and acknowledge the use of the computing facilities of Vrije Universiteit Amsterdam and SCM. Edoardo Spadetto acknowledges funding from the European Union's Horizon 2020 research and innovation program under grant agreement No 956813 (2Exciting).
\end{acknowledgement}


 \bibliography{bibo.bib,all.bib}

\begin{tocentry}
\includegraphics[width=0.8\textwidth]{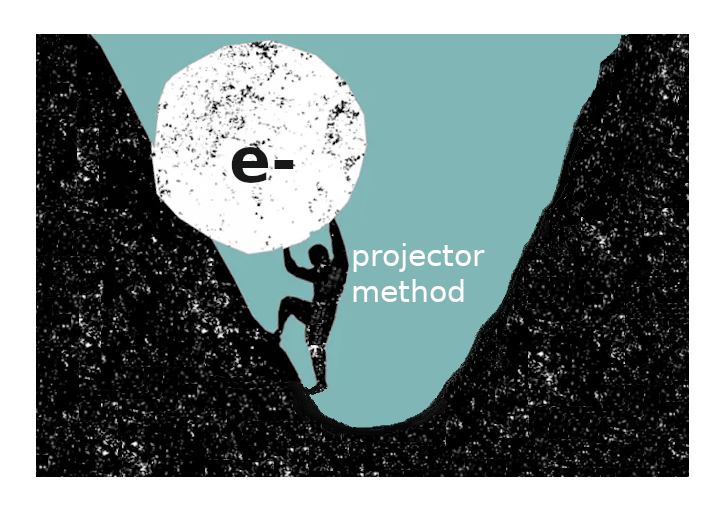}
\end{tocentry}

 \end{document}